\documentclass[epjc3]{svjour3}
\usepackage{amsmath}
\usepackage{amssymb}
\usepackage{wasysym}
\usepackage{graphicx}
\usepackage{subfigure}
\newcommand{\e}{\mbox{e}}

\journalname{Eur. Phys. J. C}
\begin{document}
\title{$p-$forms non-minimally coupled to gravity in Randall-Sundrum scenarios}
\author{G. Alencar\thanksref{e1,adr1}\and I. C. Jardim\thanksref{e2,adr2} \and R.R. Landim\thanksref{e3,adr1}}

\thankstext{e1}{e-mail: geovamaciel@gmail.com}
\thankstext{e2}{e-mail: ivan.jardim@urca.br}
\thankstext{e3}{e-mail: renan@fisica.ufc.br}
\institute{Departamento de F\'{\i}sica, Universidade Federal do Cear\'{a},
Caixa Postal 6030, Campus do Pici, 60455-760, Fortaleza, Cear\'{a}, Brazil.
\label{adr1}
\and
Universidade Regional do Carir\'{\i},
Departamento de F\'{\i}sica,
Centro de Ci\^encias e Tecnologia, Campus Crajubar.
Av. Le\~ao Sampaio, 170, Tri\^angulo
CEP: 63040-000, Juazeiro do Norte, Cear\'a.
\label{adr2}
}
\maketitle
\begin{abstract}
In this paper we study the coupling of $p$-form fields with geometrical tensor 
fields, namely Ricci, Einstein, Horndeski and Riemann in Randall-Sundrum 
scenarios with co-dimension one. We consider delta-like and  branes  generated 
by a kink and a domain wall. We begin by a detailed study of the Kalb-Ramond 
(KR) field. The analysis of KR field is very rich since it is a tensorial object 
and more complex non-minimal couplings are possible. The generalization to $p$-forms can provide more information 
about the properties and structures that can possibly be universal in the geometrical 
localization mechanism.  The zero mode is treated separately and conditions 
for localization of zero modes of $p-$forms are found for all the cases above and with 
this we arrive at the above conclusion about vector fields. Another property that can be tested 
is the absence of resonances found in the case of vector fields. For 
this we analyze the possible unstable massive modes for all the above cases via transmission coefficient. 
Our conclusion is that we have more probability to observe massive 
unstable modes in the Ricci and Riemann coupling.
\keywords{Brane-World, Localization, Resonances}
\end{abstract}

\section{Introduction}
Since Kaluza and Klein introduced extra dimensions in high energy physics to 
unify electromagnetism and gravitation\cite{Kaluza:1984ws,Klein:1926tv}, it has 
been the subject of many developments\cite{Bailin:1987jd}. In order to recover 
the 
four dimensional physics it was imposed that the extra dimension should be 
compact. However, at the end of the last century, Randall and Sundrum (RS) 
proposed an alternative to compactification using the concept of 
brane-words\cite{Randall:1999vf}. In 
this scenario the extra dimension is not compact and gravity is trapped to the 
four dimensional membrane by the introduction of a non-factorisable metric. 
Since the extra dimension is not compact all the matter fields, and not only 
gravity, must be trapped on the brane to provide a realistic model. However 
unlike gravity and the scalar field, the vector field is not trapped on the 
brane, what becomes a drawback to the RS model. To circumvent this problem some 
authors introduces a dilaton coupling\cite{Kehagias:2000au}, while others 
proposes that a strongly coupled gauge theory in five dimensions can generate a 
massless photon in the brane\cite{Dvali:1996xe}. Most of these models 
introduces other fields or nonlinearities to the gauge 
field\cite{Chumbes:2011zt}. Some 
years ago Ghoroku et al proposed a mechanism that does not includes new degrees 
of freedom and trap the gauge field to the membrane. This is based on the 
addition 
of two mass terms, one in the bulk and another on the 
brane\cite{Ghoroku:2001zu}. Despite working, the mechanism has the undesirable 
feature of possessing two free parameters, from which one is left after 
imposing the boundary conditions. Beyond this, the mechanism is not covariant 
since in principle it introduces a four dimensional mass term which does not come 
from the five dimensional bulk.

An important point about the presence of more extra dimensions is that it 
provides the existence of many antisymmetric tensor fields. In five dimensions 
for example we have the two, three, four and five forms. From the physical 
viewpoint, they are 
of great interest because they may have the status of fields describing 
particles other than the usual ones. As an example we can cite the spacetime 
torsion\cite{Mukhopadhyaya:2004cc} and the axion 
field\cite{Arvanitaki:2009fg,Svrcek:2006yi} that have separated descriptions by 
the two-form. Besides this, String Theory shows the naturalness of higher rank 
tensor fields in its spectrum\cite{Polchinski:1998rq,Polchinski:1998rr}. Other 
applications of these kind of fields have been made showing its relation with 
the AdS/CFT conjecture\cite{Germani:2004jf}. In the RS scenario much has been 
considered on these tensors. Localization of the zero mode of $p-$forms in 
delta like branes was first studied in Ref.\cite{Kaloper:2000xa} where it was 
claimed 
that, in $D$ spacetime dimensions, only forms with $p<(D-3)/2$ have a zero mode 
localized. However, it is well known that in the absence of a topological 
obstruction, the field strength of a $p-$form is Hodge dual to the 
$(D-p-2)-$form\cite{Duff:1980qv}. Using this property is was shown that in fact 
only for the $0-$form and its dual, the $(D-2)-$form, the fields are 
localized\cite{Duff:2000se,Hahn:2001ze}. Recently the authors of 
Refs\cite{Fu:2015cfa,Fu:2016vaj} showed that this is also related to the gauge 
fixing of the form fields. This make the problem of localization worse, since 
the vector field is not localized for any spacetime dimension. Beyond the zero 
mode, massive modes are important to be considered. Despite the fact that they 
are not localized, unstable massive modes can be found over the brane by using, 
for example, the transfer matrix 
method\cite{Landim:2011ki,Landim:2011ts,Alencar:2012en,Mendes:2017hmv}.  Resonances of form 
fields has been found to exist for thick and thin 
branes\cite{Alencar:2010mi,Alencar:2010hs,Alencar:2010vk,Landim:2010pq,Zhong:2010ae,Fu:2011pu,Fu:2012sa,Fu:2013ita,Du:2013bx}.  
Recently the Ghoroku mechanism was used to trapp the zero mode of $q-$form 
field\cite{Jardim:2014vba}. The point is that the introduction of the mass 
terms break the Hodge duality and the argument of 
Refs.\cite{Duff:2000se,Hahn:2001ze,Fu:2015cfa,Fu:2016vaj} is not valid anymore. 
However this solution keeps the above cited undesired 
features of the Ghoroku mechanism. 

In order to solve the above issues, recently a new proposal called "geometrical 
localization mechanism" was born\cite{Alencar:2014moa,Zhao:2014iqa}. Looking 
for a covariant version of the Ghoroku mechanism some of the present authors 
found 
that both mass terms can be obtained from a bulk action if the Ricci scalar is 
coupled to a quadratic mass term of the gauge field\cite{Alencar:2014moa}. 
Beyond solving 
the covariantization problem, it also eliminated from the beginning one of the 
free parameters. The last one is fixed by the boundary conditions leaving no 
free 
parameters in the model.  The mechanism also keep the advantage of do not 
adding any new degrees of freedom. Another good property is that it provides the 
trapping 
of the gauge fields for any smooth version of the RS 
scenario\cite{Alencar:2014moa}. Soon latter many developments of the idea was 
put forward. The same non-minimal 
coupling with the Ricci scalar was proven to work for $q-$forms and ELKO 
spinors fields\cite{Alencar:2014fga,Jardim:2014cya,Jardim:2014xla}. For 
non-abelian 
gauge fields it has been found that the non-minimal coupling with the field 
strength used in Ref.\cite{Horndeski:1976gi,Germani:2011cv} should also be 
introduced and the mechanism works\cite{Alencar:2015awa}. A phenomenological 
prevision of the 
model has been found for branes with cosmological constant different of zero: a 
precise residual mass of the photon must exist and this is proposed to be a 
probe to extra dimensions\cite{Alencar:2015rtc}. Recently the same mechanism was shown to emerge from a conformal hidden symmetry of 
the Randal-Sundrum model\cite{Alencar:2017dqb,Alencar:2017vqd}. In this new scenario fermions are shown to be universally trapped to the membrane 
by adding a non-minimal coupling with torsion. A new phenomenological prevision was found: a minimum value for the torsion of the membrane\cite{Alencar:2017dqb}. Soon 
latter a smooth version of the model was constructed in \cite{Alencar:2017vqd}. Despite providing the solution 
to the localization problem, the mechanism raises some questions. When the 
non-minimal couplings to gauge fields was generalized to includes the Ricci and the 
Einstein\cite{Alencar:2015oka}, it has been found that the last one do not 
provides a localized solution. The coupling with metric tensor also do not 
provides a 
localized zero model and this suggests that  tensors with null divergence do 
not provide a trapped gauge field. When massive modes are considered, a curious 
result is that for all smooth versions considered no resonances was found. This 
raises the question if this is an universal property of the mechanism\cite{Jardim:2015vua}.

In this paper we study the coupling of the Kalb-Ramond(KR) field with tensor 
fields. The analysis of this field is very rich since it is a tensorial 
object and more complex non-minimal couplings are possible. Beyond the above 
cited importance of the KR fields, this generalization can provide more 
information about the properties and structures that can possibly be universal 
in the geometrical localization mechanism. This paper is organized as follows. 
In the 
second section we make a briefly review the RS scenario in co-dimension one.  
In 
section III we study the localization of the zero mode  of KR field coupled to
Ricci, Einstein, Horndeski and Riemann tensors. In the section four we study 
the 
possible existence of unstable massives modes for KB field coupled  to Ricci, 
Einstein, Horndeski and Riemann tensors in a RS, kink and domain wall 
scenarios. 
In the section five we study the localization of the zero mode of the  $p-$form 
field coupled to Ricci, Einstein, Horndeski and Riemann tensors. In the six 
section we study the possible existence of unstable massive modes of the 
$p-$form field 
coupled  to Ricci, Einstein, Horndeski and Riemann tensors in a RS, kink and 
domain wall scenarios.  Finally, in the conclusions we discuss the results.

\section{Co-dimension one Randall-Sundrum Scenario}

Due to the variety of geometrical objects needed in this manuscript, in this 
section we briefly review the RS scenario in co-dimension one
brane world in $D$-dimensional space-time and construct explicitly all the geometric tensors needed. The coordinates 
of the whole space-time are $x^M$, $M=0,1,2,\cdots D-1$  with $x^{D-1}\equiv z$ the coordinate transverse to the 
brane and $x^\mu, \mu=0,1,2,\cdots,D-2$ is the usual Minkowski  coordinates. The 
metric is  $ds^{2}=\e^{2A(z)}\eta_{MN}dx^{M}dx^{N}$ where $\eta_{MN}=\mbox{diag}(-++\cdots+)$ and
the equations of motion are given by \cite{Randall:1999vf}
\[
\sqrt{-G}\left(R_{MN}-\frac{1}{2}G_{MN}R\right)=-\frac{1}{4M^{3}}
\left(\Lambda\sqrt{-G}G_{MN}+V\sqrt{-g}g_{\mu\nu}\delta_{M}^{\mu}\delta_{N}^{\nu
}\delta(z)\right),
\]
where $\Lambda$ is the cosmological constant and $V$ is the brane tension. The 
conformal form of the metrics provides a simple way to obtain
the needed geometrical quantities. Interestingly this will also provide
a covariant description of the model. First we must remember that,
under a conformal transformation $\tilde{g}_{MN}=\e^{2\varphi}g_{MN}$,
we have for the Christoffel symbols
\[
\tilde{\Gamma}^{K}{}_{IJ}=\Gamma^{K}{}_{IJ}+\delta_{I}^{K}\partial_{J}
\varphi+\delta_{J}^{K}\partial_{I}\varphi-g_{IJ}\partial^{K}\varphi.
\]
The transformation of the Ricci tensor and Ricci scalar are 
\begin{eqnarray} 
\tilde{R}_{IJ}&=&R_{IJ}-(D-2)\left[\nabla_{I}\partial_{J}\varphi-\partial_{I}
\varphi\partial_{J}\varphi\right]-\left(\square\varphi+(D-2)\|\nabla\varphi\|^{2
}\right)g_{IJ};\nonumber \\
\tilde{R}&=&\e^{-2\varphi}\left(R-2(D-1)\square\varphi 
-(D-2)(D-1)\|\nabla\varphi\|^{2}\right),\nonumber 
\end{eqnarray}
where $\nabla_{I}$ is the covariant derivative. From these we can get the 
transformation of the Einstein tensor, given by
\begin{equation} 
\tilde{G}_{IJ}=G_{IJ}-(D-2)\left[\nabla_{I}\partial_{J}\varphi-\partial_{I}
\varphi\partial_{J}\varphi \right]+(D-2)\left[ \square\varphi 
+\frac{(D-3)}{2}\|\nabla\varphi\|^{2}\right]g_{IJ}.
\end{equation}

In the RS case we have $g_{MN}=\eta_{MN}$ and $\varphi=A(z)$ and
this gives us for the components of the Christoffel symbols
\[
\Gamma_{D-1\,D-1}^{D-1}=A',\Gamma_{\mu\nu}^{D-1}=-A'\eta_{\mu\nu,}\Gamma_{D-1\,\alpha}^{\mu}
=\delta_{\alpha}^{\mu}A'.
\]
For the components of the Ricci scalar, Ricci and Einstein tensors we have
\begin{eqnarray}
&&R=-(D-1)\e^{-2A}(2A''+(D-2)A'^{2}),\nonumber\\
&&R_{D-1\,D-1}=-(D-1)A'', R_{\mu\nu}=-\eta_{\mu\nu}(A''+(D-2)A'^{2}),\nonumber\\
&&G_{D-1\,D-1}=(D-2)(D-1)A'^2/2,G_{\mu\nu}=(D-2)(A''+\frac{(D-3)}{2}A'^{2})\eta_{
\mu\nu}.
\end{eqnarray}

With the above results we get for the Einstein equations 
\begin{eqnarray}
(D-2)A''+\frac{1}{2}(D-2)(D-3)A'^{2}&=&-\frac{V}{4M^{3}}\delta(z)-\frac{\Lambda}
{4M^{3}}\e^{2A}\nonumber\\
\frac{1}{2}(D-1)(D-2)A'^{2}&=&-\frac{\Lambda}{4M^{3}}\e^{2A}\label{EE}
\end{eqnarray}
with solution
\[
A(z)=-\ln(k|z|+1),V=8M^{3}k(D-2),\Lambda=-2M^{3}(D-1)(D-2)k^{2}.
\]
Therefore we see that the solution is identical as in the five dimensional case 
but with the tension of the brane and the cosmological constant
depending on the spacetime dimension.

It is important to point that in this manuscript, non-minimal couplings with 
quadratic higher order antisymmetric tensors will be considered. Therefore we 
must look for
higher order geometric tensors which has the same symmetries. The first 
geometric tensor of order four with this properties is the Riemann tensor. 
Under a conformal transformation it changes to
\[
\tilde{R}_{IJKL}=\e^{2\varphi}\left(R_{IJKL}-\left[
g\odot\left(\nabla\partial\varphi-\partial\varphi\partial\varphi+\frac{1}{2}
\|\nabla\varphi\|^{2}g\right)\right]_{IJKL}\right),
\]
where $\odot$ is the Kulkarni\textendash Nomizu product defined by
\[
\left(h\odot 
k\right)_{IJKL}=h_{IK}k_{JL}+h_{JL}k_{IK}-h_{IL}k_{JK}-h_{JK}k_{IL}.
\]
When considering the RS metrics the components are simplified to
\begin{equation}\label{Riemann}
R_{\mu\nu\alpha\,D-1}=0,R_{\mu\,D-1\nu\,D-1}=-\eta_{\mu\nu}(A''+2A'^{2})\e^{2A},R_{
\mu\nu\alpha\beta}=-A'^{2}\e^{2A}(\eta_{\mu\alpha}\eta_{\nu\beta}-\eta_{
\nu\alpha}\eta_{\mu\beta}).
\end{equation}

However, the curvature tensor has non-null divergence. As said in the 
introduction, we must also analyze tensors with null divergence. A fourth order 
tensor with this 
property is the Horndeski tensor and has been coupled to the field strength of 
the vector field Ref.\cite{Horndeski:1976gi,Germani:2011cv,Alencar:2015awa}. 
Curiously this 
tensor has all the desired symmetries. Here we will consider the coupling this 
tensor to the mass term of the form field. It is given by
\[
\Delta^{AB}{}_{CD}\equiv\frac{1}{8}R^{AB}{}_{CD}-\frac{1}{2}R^{[A}{}_{[C}\delta^
{B]}{}_{D]}+\frac{1}{8}R\delta^{[A}{}_{[C}\delta^{B]}{}_{D]},
\]
where
$$
M^{[A}{}_{[C}N^{B]}{}_{D]}\equiv\frac{1}{4}(M^{A}_{C}N^{B}_{D}-M^{A}_{D}N^{B}_{C
}+M^{B}_{D}N^{A}_{C}-M^{B}_{C}N^{A}_{D}).
$$
We should point that, since it has null divergence in any index, after 
contracting two of them we must obtain a tensor proportional to the Einstein 
tensor. In fact, by a direct calculation we get
$$
\Delta^{AB}{}_{CB}=-\frac{D-3}{8}G^{A}_{C}.
$$

Under a conformal transformation we have
\[
\tilde{\Delta}^{AB}{}_{CD}=\Delta^{AB}{}_{CD}-\e^{-2\varphi}(D-3)\mathcal{H}^{AB}_{CD},
\]
where
\[
 \mathcal{H}^{AB}_{CD}=\left(\frac{1}
{4}\square \varphi+\frac{(D-4)}{8}(\nabla 
\varphi)^{2}\right)\delta^{[A}{}_{[C}\delta^{B]}{}_{D]}-\frac{1}{2}\left(\delta_{[C}^{[A}
\partial^{B]}\partial_{D]}\varphi-\delta_{[C}^{[A}\partial^{B]}\varphi\partial_{
D]}\varphi\right).
\]
Using the RS metrics we obtain the components
\[
\Delta^{\mu\nu}{}_{\alpha\beta}=-\frac{\e^{-2A}}{8}(D-3)(\delta_{\alpha}^{\mu}
\delta_{\beta}^{\nu}-\delta_{\beta}^{\mu}\delta_{\alpha}^{\nu})(A''+\frac{(D-4)}
{2}A'^{2}),
\]
and

\[
\Delta_{\nu\,D-1}^{\mu\,D-1}=-\frac{\e^{-2A}}{16}\delta_{\nu}^{\mu}(D-3)(D-2)A'^{2}.
\]

We will use the above results to study a variety of geometrical couplings
which can renders localized modes for the fields.
 First we will study zero mode localization and then the massive modes. In the section 3 we will restrict our analysis to 
 five dimensional case.

\section{The Kalb-Ramond zero mode case}
In this section we will make a direct generalization of the geometric coupling, 
presented in Ref.\cite{Alencar:2015oka}, for the KR field. This 
is gonna be a prototype for the due generalization to the $q-$form field in the 
section five. Beyond this, due to its importance it is worthwhile to make 
a separate study. We will consider the coupling to tensors of order two and 
four.

\subsection{Kalb-Ramond coupled with a rank two geometric tensor}
In this subsection we will consider the coupling of the KR field to rank two 
geometric tensors. 
The action is given by 
\begin{equation}\label{SKR2}
 S = -\int 
d^{5}x\sqrt{-g}\left[\frac{1}{12}Y_{M_{1}M_{2}M_{3}}Y^{M_{1}M_{2}M_{3}} 
+\frac{1}{2}\gamma 
g_{N_{1}N_{2}}H_{M_{1}M_{2}}X^{M_{1}N_{1}}X^{M_{2}N_{2}}\right],
\end{equation}
where $X_{MN}$ is the anti-symmetric Kalb-Ramond two form field and 
$Y_{MNO}=\partial_{[M}X_{NO]}$. We also use $H_{M_{1}M_{2}}$ as a generic rank 
two geometric tensor and $\gamma$ is a coupling constant. The above action 
provides the following equation of motion for the KR field
\begin{equation}\label{H2full}
 \frac{1}{2}\partial_{M_{1}}\left[\sqrt{-g}Y^{M_{1}M_{2}M_{3}}\right] 
-\gamma\sqrt{-g} g^{N_{1}[M_{3}}H^{M_{2}]M_{1}}X_{M_{1}N_{1}} = 0,
\end{equation}
and from this, we get the identity
\begin{equation}\label{H2identityfull}
\partial_{M_{2}}\left[\sqrt{-g}g^{N_{1}[M_{3}}H^{M_{2}]M_{1}}X_{M_{1}N_{1}}
\right] = 0.
\end{equation}

The analyzes can be simplified if we observe that in all the cases the tensors 
has the same form, namely
\begin{equation}\label{H2}
H_{\mu\nu}=H_0\eta_{\mu\nu}, H_{44}=H_1,
\end{equation}
and all other components are null. 
For a free index equal to extra dimension index, i.e., $M_{3} = 4$ we obtain 
from the equations of motion (\ref{H2full}) the vectorial equation
\begin{equation}\label{eqKR25}
 \partial_{\mu}Y^{\mu\nu4} -\gamma(H_{0}+H_{1})X^{\nu4} = 0,
\end{equation}
where  in the above expression and from now on all the indexes are raised with 
$\eta^{\mu\nu}$. Taking $M_{3} = \mu_{3}$ in (\ref{H2full}) we obtain a 
tensorial equation
\begin{equation}
 \frac{1}{2}\e^{-A}\partial_{\mu_{1}}Y^{\mu_{1}\mu_{2}\mu_{3}} + 
\frac{1}{2}\partial_z\left[\e^{-A}Y^{4\mu_{2}\mu_{3}}\right] -\gamma\e^{-A} 
{H}_{0}X^{\mu_{2}\mu_{3}} = 0.
\end{equation}

Taking the free index $M_{3} = 4$ in the identity (\ref{H2identityfull}), we 
obtain the gauge condition for the vector field, i.e., 
$\partial_{\mu_{1}}X^{\mu_{1}4} = 0$.
For the free index $M_{3} = \mu_{3}$ we obtain
a tensorial condition
\begin{equation}\label{divH2}
\e^{-A}{H}_{0}\partial_{\mu_{1}}X^{\mu_{1}\nu_{1}} 
+\frac{1}{2}\partial_z\left[\e^{-A}({H}_{1} +{H}_{0})X^{4\nu_{1}}\right]= 0.
\end{equation}
 Therefore the KR field is not divergence free and we must decompose it as 
$X^{\mu_{1}\mu_{2}}=X_{L}^{\mu_{1}\mu_{2}}+X_{T}^{\mu_{1}\mu_{2}}$, with
\begin{equation}\label{longKR}
X_{T}^{\mu_{1}\mu_{2}}=X^{\mu_{1}\mu_{2}}+\frac{1}{\Box}\partial^{[\mu_{1}}
\partial_{\nu_{1}}X^{\mu_{2}]\nu_{1}};\;\;\;\;X_{L}^{\mu_{1}\mu_{2}}=-\frac{1}{
\Box}\partial^{[\mu_{1}}\partial_{\nu_{1}}X^{\mu_{2}]\nu_{1}}.
\end{equation}
where $X_{T}^{\mu_{1}\mu_{2}}$ has null divergence, as desired. From now on, we 
must show that the transversal and longitudinal parts of the field 
decouples in the equations of motion. For this, we first show that from the 
above 
definitions we get
\begin{equation}\label{YLong}
 \partial_{\mu_{1}}Y^{\mu_{1}\mu_{2}\mu_{3}}=2\square 
X_{T}^{\mu_{2}\mu_{3}};\;\;\;\; Y^{\mu_{1}\mu_{2}4}= Y^{\mu_{1}\mu_{2}4}_{L} 
+2\partial X^{\mu_{1}\mu_{2}}_{T},
\end{equation}
where 
$Y^{\mu_{1}\mu_{2}4}_{L}\equiv\partial^{\mu_1}X^{\mu_24}-\partial^{\mu_2}X^{
\mu_14}$. Using the above result  in the field equations we arrive at 
\begin{eqnarray}
&& \e^{-A}\Box X^{\mu_{2}\mu_{3}}_{T} +\partial_z\left[\e^{-A}\partial_z 
X^{\mu_{2}\mu_{3}}_{T}\right] + 
\frac{1}{2}\partial_z\left[\e^{-A}Y^{\mu_{2}\mu_{3}4}_{L} \right]\nonumber\\ 
&&-\gamma\e^{-A}{H}_{0}X^{\mu_{2}\mu_{3}}_{L} - \gamma\e^{-A} 
{H}_{0}X^{\mu_{2}\mu_{3}}_{T} = 0, \\ \label{KR2coup1}
&& \partial_{\mu}Y^{\mu\nu4}_{L} -\gamma({H}_{0}+ {H}_{1})X^{\nu4} = 0 
\label{KR2coup2}.
\end{eqnarray}

From the definition of $Y^{\mu_{1}\mu_{2}4}_{L}$ we can also show that
\begin{equation}\label{Yinverse}
 Y^{\mu_{2}\mu_{3}4}_{L} = 
-\frac{1}{\Box}\partial^{[\mu_{2}}\partial_{\nu}Y^{\mu_{3}]\nu4},
\end{equation}
and using Eqs.(\ref{eqKR25}), (\ref{divH2}) and (\ref{longKR}) we can finally 
show that the longitudinal propagator term that appear in (\ref{KR2coup1}) is 
equal to
\begin{eqnarray}
  \partial_z\left[\e^{-A}Y^{\mu_{2}\mu_{3}4}_{L} \right] =  
\gamma\partial_z\left[\e^{-A}({H}_{0} 
+{H}_{1})\frac{1}{\Box}\partial^{[\mu_{2}}X^{\mu_{3}]4} \right]\nonumber\\ 
=  -2\gamma 
H_{0}\e^{-A}\frac{1}{\Box}\partial^{[\mu_{2}}\partial_{\mu_{1}}X^{\mu_{3}]\mu_{1}
} = 2\gamma H_{0}\e^{-A}X_{L}^{\mu_{2}\mu_{3}}.
\end{eqnarray}
The above identity can be used to decouples the fields in Eq. (\ref{KR2coup1}), 
and provides the final equation for the transversal component 
\begin{equation}
  \e^{-A}\Box X^{\mu_{2}\mu_{3}}_{T} +\partial_z\left[\e^{-A}\partial_z 
X^{\mu_{2}\mu_{3}}_{T}\right]  -\gamma\e^{-A} {H}_{0}X^{\mu_{2}\mu_{3}}_{T} = 0.
\end{equation}

Performing the separation of variables in the form $ X_{T}^{\mu_{1}\mu_{2}} = 
\tilde{X}_{T}^{\mu_{1}\mu_{2}}(x)\e^{A/2}\psi_{T}(z)$, we obtain the set of 
equations
\begin{eqnarray}
&&  \Box \tilde{X}^{\mu_{2}\mu_{3}}_{T} 
-m^{2}_{T}\tilde{X}^{\mu_{2}\mu_{3}}_{T} 
= 0,
\\&& \psi_{T}''-U_{T}(z)\psi_{T}  = -m^{2}_{T}\psi_{T}, \label{eqKR2}
 \end{eqnarray}
where the potential is given by 
\begin{equation}\label{potKR2}
 U_{T}(z) = \frac{A'^{2}}{4} -\frac{A''}{2}  + \gamma{H}_{0}.
\end{equation}

To decouple the vector field and the longitudinal part of KR field we use the 
divergence equation (\ref{divH2}) in (\ref{KR2coup2}). This procedure leads to
\begin{equation}\label{vecH2}
\Box X^{\nu4} +\partial_z\left[\frac{\e^{A}}{2H_{0}}\partial_z\left(\e^{-A}(H_{0} 
+H_{1})X^{\nu4}\right)\right] -\frac{\gamma}{2}(H_{0}+H_{1})X^{\nu4} = 0.
\end{equation}
To write the above equation in a Schr\"odinger-like form we must separate the 
variables in the form $X^{\nu4} = \tilde{X}^{\nu}(x)F(z)\psi(z)$, where
\begin{equation}\label{F}
 F(z) = \frac{2\e^{A/2}({H}_{0})^{1/2}}{{H}_{0}+{H}_{1}}.
\end{equation}
This procedure splits the equation (\ref{vecH2}) in the following set of 
equations
\begin{eqnarray}
&& \Box\tilde{X}^{\nu} - m^{2}\tilde{X}^{\nu} = 0
\\&& \psi''-U(z)\psi  = -2m^{2}\frac{{H}_{0}}{{H}_{0} +{H}_{1}}\psi, 
\label{eqvec2}
\end{eqnarray}
where the potential of Schr\"odinger equation is given by
\begin{equation}\label{potvec2}
 U(z) = \frac{1}{4}\left[A' -(\ln{H}_{0})'\right]^{2} +\frac{1}{2}\left[A 
-\ln{H}_{0}\right]'' +\gamma{H}_{0}.
\end{equation}
The equations (\ref{eqKR2}) and (\ref{eqvec2}), with the potentials 
(\ref{potKR2}) and (\ref{potvec2}), governs the extra dimension component of KR 
and vector 
fields respectively. 

Now we must analyze the localization of the zero mode of the KR and vector 
fields. We first consider the KR case. For this we must to explicit the 
geometric tensor which couples with KR field in the action (\ref{SKR2}). We 
will 
see that this can be achieved without any specific form for the warp factor. 
The 
only condition is that the RS model is recovered for large $z$. First of all, 
the potential (\ref{potKR2}) can further be simplified if we note that $H_{0}$ 
are combinations of $A'^{2}$ and $A''$ 
\begin{equation}\label{2H_0H_1}
H_{0} = \lambda_0 A'' +\beta_0 A'^{2},H_{1} = \lambda_1 A'' +\beta_1 A'^{2},
\end{equation}
what gives us its final form
\begin{equation}
 U_{T}(z) = (\frac{1}{4}+\gamma\beta_0)A'^{2} -(\frac{1}{2}-\gamma\lambda_0)A''.
\end{equation}

Supposing now the solution for the zero mode $\psi(z) = \e^{\sigma A(z)}$, we 
obtain from Eq.(\ref{eqKR2}) with $m=0$ the set of algebraic equations 
\begin{eqnarray}
&& \gamma(-\gamma\lambda_0^2+\lambda_0+\beta_0) =0,\\
&& \sigma = -(1/2 - \gamma\lambda_0).
\end{eqnarray}
The solution $\gamma=0$ is the solution for the free KR field and gives us a non 
localized field as expected. For $\lambda_0=0$ we have $\gamma\beta_0=0$ what 
also 
implies the free field solution which is not localized. The last possibility is 
$\gamma\neq 0$ and $\lambda_0\neq 0$. With this we get the solutions
\begin{eqnarray}\label{gammaH2}
&& \gamma=\frac{\lambda_0+\beta_0}{\lambda_0^2},\\
&& \sigma = 1/2 + \frac{\beta_0}{\lambda_0}.
\end{eqnarray}

Therefore to obtain a convergent solution for the zero mode of the transversal 
part of KR field, a necessary condition is that $\beta_0/\lambda_0>0$. This is 
valid 
in all brane scenarios that recover the RS asymptotically.

The first example is the Ricci tensor. From the last section we can see that
$$
 H_{0}= -(A''+3A'^{2}), H_{1} = -4A'',
$$
and therefore $\lambda_0=-1,\beta_0 =-3$, which gives us a localized zero mode 
if 
$\gamma = -4$ with $\sigma=7/2$.  Therefore, just as in the case for the vector 
field \cite{Alencar:2015oka}, the KR field we can be localized with the Ricci 
tensor. The Einstein tensor is the second example of a rank two geometric 
tensor. From the 
last section we can see that in this case
\begin{equation}
H_{0} =  3(A'' +A'^{2}), H_{1} =  6A'^{2},
\end{equation}
and we have $\lambda_0=3,\beta_0 =3$. This value provides a localized solution 
for the zero mode of transversal part of KR field for a coupling constant 
fixed by $\gamma = 2/3$ with $\sigma=3/2$. 

Next we will analyze the localization of the zero mode of the vector component 
of the KR field. In this case, due to the transformation (\ref{F}), the 
integral 
that must be finite is given by
\begin{equation}\label{integrand}
\int \frac{H_{0}}{(H_{0}+H_{1})^2}\psi^2dz,
\end{equation}
where $\psi$ is the solution of (\ref{eqvec2}) with $m=0$. However, different 
of 
the tensor case, it is not possible to find analytical solutions of 
Eq.(\ref{eqvec2}). 
However, a necessary condition for localizability is a convergent solution for 
large $z$. Since all the smooth versions considered here recover RS for large 
$z$, we can 
use this solution to test the localizability of the field. With these 
considerations we can use the Einstein equation (\ref{EE}), which will be valid 
for 
large $z$, to obtain that 
\begin{equation}\label{asymptotic}
A''=A'^2=k^2e^{2A},
\end{equation}
and therefore
\begin{equation}\label{Hasymptotic}
H_0=(\lambda_0+\beta_0)k^2e^{2A}, H_1=(\lambda_1+\beta_1)k^2e^{2A}.
\end{equation}

From the above relation we get for the potential (\ref{potvec2}) at large $z$
\begin{equation}\label{efectivepotvec2}
 U(z) = \frac{A'^2}{4}-\frac{A''}{2}+\gamma H_0.
\end{equation}
This is the same potential found for the KR case, Eq.(\ref{potKR2}). Therefore 
the solution is the same, namely $\psi=e^{\sigma A}$, with the 
same $\sigma$ and $\gamma$ as before. However, since the integrand for the 
vector case is given by Eq.(\ref{integrand}) the condition for localizability 
is different. In the limit considered here we can use  (\ref{Hasymptotic}) and 
the integrand of Eq.(\ref{integrand}) reduces to $ e^{2A}\psi^2$. Therefore the 
condition for localization in this case is given by $\beta_0/\lambda_0>-1$. 
By this fact, we can conclude that  for any case in which the KR is localized we also 
have a localized vector field. With this, we show that the hypothesis of 
Ref.\cite{Alencar:2015oka} that tensors with null divergence do not trap zero 
modes is no valid. 
This is true since the Einstein tensor can localize the KR field. 

\subsection{Kalb-Ramond coupled with a rank four geometric tensor}
In this section we will extend the coupling used in last subsection to a rank 
four geometric tensor. As said in the first section these tensors must have 
the same symmetries of the Riemann tensor in order to couple with the quadratic 
KR field. The action is given by
\begin{equation}\label{SKR4}
 S = -\int 
d^{5}x\sqrt{-g}\left[\frac{1}{24}Y_{M_{1}M_{2}M_{3}}Y^{M_{1}M_{2}M_{3}} 
+\frac{1}{4}\gamma H_{M_{1}N_{1}M_{2}N_{2}}X^{M_{1}N_{1}}X^{M_{2}N_{2}}\right],
\end{equation}
where $H_{M_{1}N_{1}M_{2}N_{2}}$ is a generic geometric tensor. The equations 
of 
motion are
\begin{equation}\label{H4full}
 \frac{1}{2}\partial_{M_{1}}\left[\sqrt{-g}Y^{M_{1}M_{2}M_{3}}\right] 
-\gamma\sqrt{-g} H^{M_{1}N_{1}M_{2}M_{3}}X_{M_{1}N_{1}} = 0,
\end{equation}
and from this we get the constraint
\begin{equation}\label{H4identityfull}
\partial_{M2}(\sqrt{-g} H^{M_{1}N_{1}M_{2}M_{3}}X_{M_{1}N_{1}}) = 0.
\end{equation}

Now we proceed to decompose these equations in components. As previously 
pointed, before this we should note that all tensors considered by us have the 
following structure
\begin{eqnarray}\label{H4}
&& H^{\mu_{1}\mu_{2}\mu_{3}\mu_{4}} = 
\frac{1}{2}\e^{-6A}H_{0}(\eta^{\mu_1\mu_3}\eta^{\mu_2\mu_4}-\eta^{\mu_2\mu_3}
\eta^{\mu_1\mu_4}), \nonumber
\\&& H^{\mu_{1}4\mu_{2}4}=\frac{1}{2}\e^{-6A}H_{1}\eta^{\mu_{1}\mu_{2}}, 
\nonumber
\\&& H^{\mu_{1}\nu_{1}4\mu_{3}} =H^{\mu_{1}4\mu_{2}\mu_{3}}= 0.
\end{eqnarray}
With this, we obtain from Eq.(\ref{H4full}) with $M_2=\mu_2,M_3=\mu_3$ and 
$M_3=4$ respectively
\begin{eqnarray}
&&\frac{1}{2}\e^{-A}\partial_{\mu_{1}}\left[Y^{\mu_{1}\mu_{2}\mu_{3}}\right] + 
\frac{1}{2}\partial_z\left[\e^{-A}Y^{4\mu_{2}\mu_{3}}\right] -\gamma H_0\e^{-A} 
X^{\mu_{2}\mu_{3}} = 0,\label{H4mu}
\\&& \frac{1}{2} \partial_{\mu}Y^{\mu\nu4} -\gamma H_1X^{\nu4}= 0,\label{H45}
\end{eqnarray}
and for the constraint (\ref{H4identityfull}) we get the following components
\begin{eqnarray}\label{H4identitycomp}
&& \partial_z\left(\e^{-A}H_1X^{4\mu_{3}} 
\right)+\e^{-A}H_0\partial_{\mu_{2}}X^{\mu_{2}\mu_{3}} = 0,
\\&& \partial_{\mu_{2}}X^{\mu_{2}4} = 0. 
\end{eqnarray}

As expected the KR field does not has null divergence and now we proceed to 
show 
that its longitudinal and transversal pieces, as defined in Eq.(\ref{longKR}), 
decouples. 
For this we use the identities (\ref{YLong}) in the equations of motion to 
obtain
\begin{eqnarray}\label{YmuL}
&&\e^{-A}\Box X^{\mu_{2}\mu_{3}}_{T} +\partial_z\left(\e^{-A}\partial_z 
X^{\mu_{2}\mu_{3}}_{T}\right) + 
\frac{1}{2}\partial_z\left(\e^{-A}Y^{\mu_{2}\mu_{3}4}_{L} \right)\nonumber\\
&&-\gamma H_0\e^{-A} X^{\mu_{2}\mu_{3}}_L-\gamma H_0\e^{-A} 
X^{\mu_{2}\mu_{3}}_T 
= 0,\\
&& \frac{1}{2} \partial_{\mu}Y^{\mu\nu4}_L -\gamma H_1X^{\nu4}= 0\label{Y5L}.
\end{eqnarray}
Now, by using the identity (\ref{Yinverse}) and Eqs.(\ref{H45}),(\ref{H4identitycomp}) and (\ref{longKR}) we can show that
$$
\partial_z\left(\e^{-A}Y^{4\mu_{2}\mu_{3}}\right)=2\gamma H_0\e^{-A} 
X^{\mu_{2}\mu_{3}}_L,
$$
and the longitudinal contribution decouples in Eq.(\ref{YmuL}). To decouple 
the vector component we just use Eq.(\ref{H4identitycomp}) in (\ref{Y5L}). With 
this, we finally get for the decoupled equations of motion
\begin{eqnarray}
&&\e^{-A}\Box X^{\mu_{2}\mu_{3}}_{T} +\partial_z\left(\e^{-A}\partial_z 
X^{\mu_{2}\mu_{3}}_{T}\right) -\gamma H_0\e^{-A} X^{\mu_{2}\mu_{3}}_T = 0,
\\&& \Box X^{\nu4} 
+\partial_z\left[\frac{\e^{-A}}{H_{0}}\partial_z\left(\e^{A}H_{1}X^{\nu4}
\right)\right]-\gamma H_{1}\e^{2A}X^{\nu4} = 0.
\end{eqnarray}

For the transversal part of KR field, making the separation  
$X^{\mu_{2}\mu_{3}}_{T} = 
\tilde{X}^{\mu_{2}\mu_{3}}_{T}(x)\e^{A/2}\psi_{T}(z)$, 
we obtain
\begin{eqnarray}
&& \Box \tilde{X}^{\mu_{2}\mu_{3}}_{T} = m^{2}_{T} 
\tilde{X}^{\mu_{2}\mu_{3}}_{T},
\\&& \psi_{T}'' -U_{T}(z)\psi_{T} = -m^{2}_{T}\psi_{T} \label{eqKR4},
\end{eqnarray}
where the Schr\"odinger's potential is given by
\begin{equation}
 U_{T}(z) =  A'^{2}/4 -A''/2 +\gamma H_{0}. \label{potKR4}
\end{equation}

To write the vector equation in a Schr\"odinger-like form we must separate the 
variables $X^{\nu4} = \tilde{X}^{\nu}(x)F(z)\psi(z)$, where
\begin{equation}\label{F4}
 F(z) = \frac{\e^{A/2}({H}_{0})^{1/2}}{{H}_{1}},
\end{equation}
and we get 
\begin{eqnarray}
&& \Box\tilde{X}^{\nu} - m^{2}\tilde{X}^{\nu} = 0,
\\&& \psi''-U(z)\psi  = -m^{2}\frac{{H}_{0}}{{H}_{1}}\psi, \label{eqvec4}
\end{eqnarray}
where the potential of Schr\"odinger equation is given by
\begin{equation}\label{potvec4}
 U(z) = \frac{1}{4}\left[A' -(\ln{H}_{0})'\right]^{2} +\frac{1}{2}\left[A 
-\ln{H}_{0}\right]'' +\gamma{H}_{0}.
\end{equation}

At this point we must analyze the localization of the zero modes of the fields. 
However we can see that the form of the equations that governs the zero 
modes are identical and in the last section. Therefore, if H decomposes as in 
(\ref{2H_0H_1}) we will have that the tensor component is 
localized for $\lambda_0\neq 0$ and $\beta_0/\lambda_0> 0$, with $\gamma$ given 
by Eq. (\ref{gammaH2}). For the Riemann tensor, 
by comparing Eq. (\ref{Riemann}) with Eqs. (\ref{H4}) and (\ref{2H_0H_1}), we 
have
\begin{equation}
H_{0} =  -2A'^{2}, H_{1} = -2(A''+2A'^{2}),
\end{equation}
and therefore $\lambda_0=0$. The conclusion is that the Riemann tensor do not 
trap the KR field. For the Horndeski tensor we have
\begin{equation}
H_{0} =  -\frac{1}{2}(A''+\frac{1}{2}A'^{2}), H_{1} = -\frac{3}{8}A'^{2},
\end{equation}
and $\lambda_0=-1/2,\beta_0=-1/4$, what gives a localized zero mode with 
$\gamma=-3$ and $\sigma=1$. Just as before, the vector field is localized 
always that the KR field is localized. At this point is curious to see that 
relation with null divergence and localization seems to be inverted. The 
Horndeski 
tensor has null divergence and provides a trapped field, while the Riemann 
tensor do not. In order to obtain a final answer for this questions we have to 
generalize 
our results to $p-$form fields. We do this in the next sections, but before we 
analyze the possible resonances for the cases considered here.

\section{Kalb-Ramond massive modes}
In this section we study the possible resonant modes with Kalb-Ramond field 
coupled to  Ricci, Einstein, Horndeski and Riemann tensors, through the 
transmission coefficient.
The resonant modes appears when the transmission 
coefficient $T$ is equal to 1, i.e, $Log(T)=0$.  We analyze in three possible 
scenarios: Randall-Sundrum delta like brane, a brane generated by a domain wall and generated by a kink.

We first observe that for all tensor coupling, the potential of Schr\"odinger equation in conformal metric has the form
\begin{equation}\label{pot-all-KR}
U_T(z)=\alpha A'(z)^2+\beta A''(z),
\end{equation}
where $\beta=\sigma=7/2,3/2,1$, $\alpha=\sigma^2$ for Ricci, Einstein and Horndeski tensors respectively and $\alpha=1/4-2\gamma$, $\beta=-1/2$ for the Riemann tensor.

\subsection{In Randall-Sundrum delta like scenario}
The first brane scenario that we will study is the Randall-Sundrum scenario 
\cite{Randall:1999vf}. Despite the singularity this scenario has a historical 
importance and serves as an important paradigm in physics of extra dimensions and field 
localization. The warp factor of this scenario in a conformal form is given by
\begin{equation}\label{WF-delta}
A(z) = -\ln\left(k|z| +1\right).
\end{equation}

\begin{figure}[!h]
\centering
\includegraphics[width=5cm]{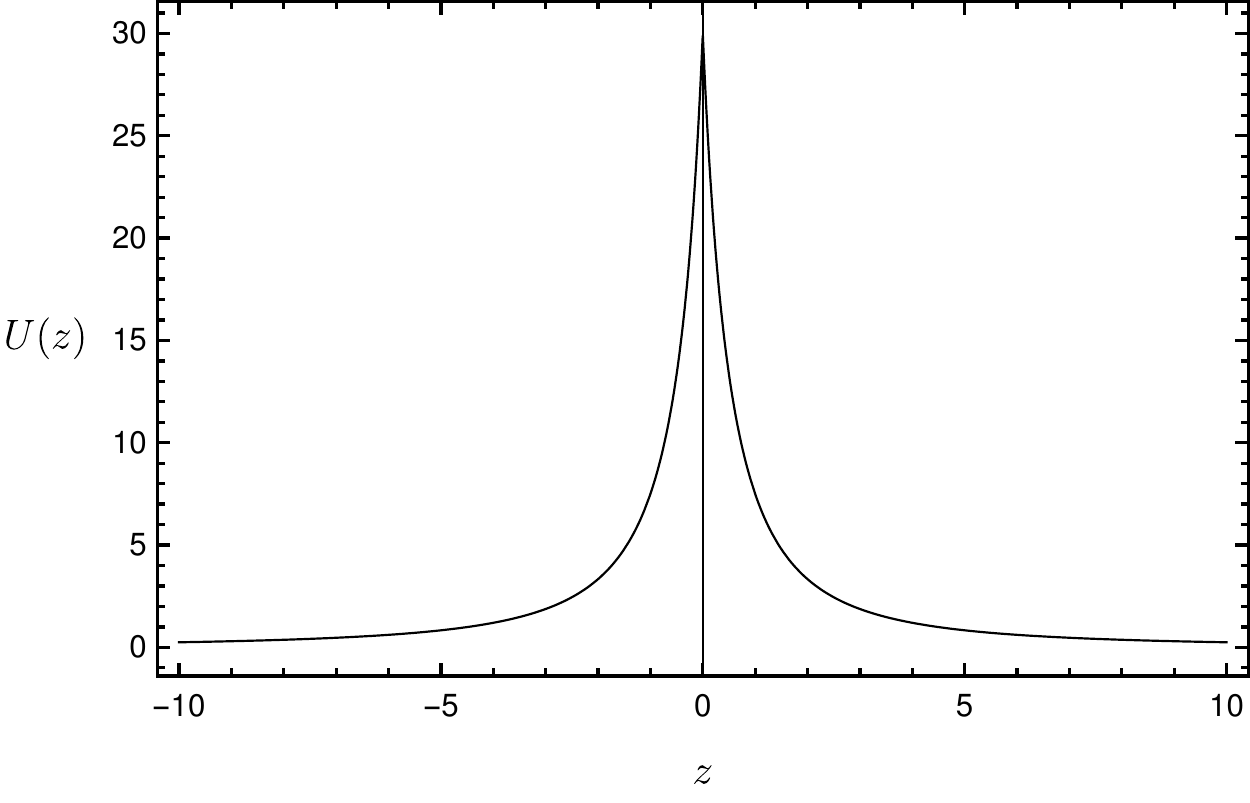}
\caption{Plot of regular part of Schr\"odinger potential for KR field in 
Randall-Sundrum delta like scenario with $k =1$ and $\alpha+\beta=30$.}
\label{fig:Profile-alpha-beta-RS}
\end{figure}

In this scenario, the potential of Schr\"odinger equation is given by 
\begin{equation}\label{potR2RS}
U_T(z)=\frac{(\alpha+\beta)k^{2}}{(k|z|+1)^{2}} -2k\beta\delta(z),
\end{equation}
and is illustrated in Fig. \ref{fig:Profile-alpha-beta-RS}. The regular part of $U_T(z)$ has a maximum  at $z=0$. For Riemann tensor $\alpha+\beta=-1/4-2\gamma$. 
We must have $\gamma <-1/8$, in order to provide a positive maximum for the potential and a positive asymptotic behavior.

\begin{figure}[!h]
\centering
\subfigure[]{
\includegraphics[width=5cm]{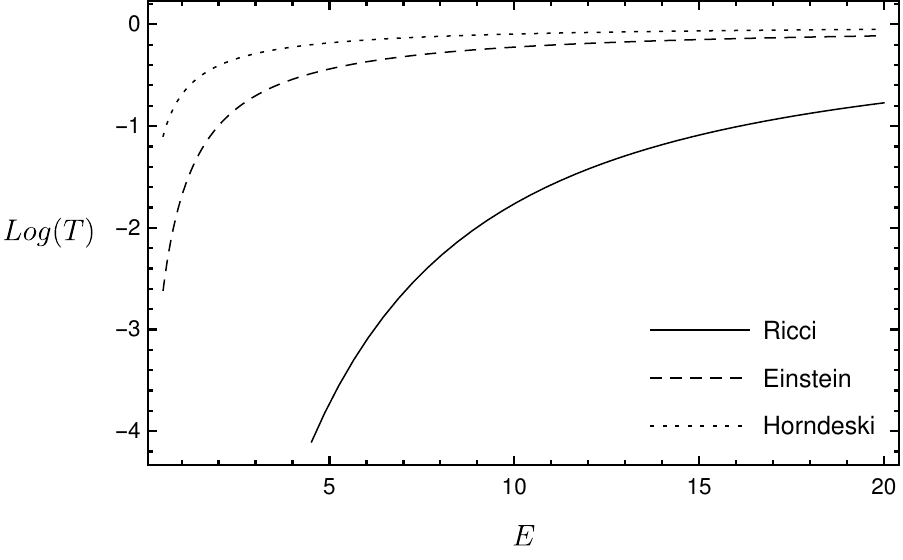}
\label{fig:TREH-RS-D5p2}
}
\subfigure[]{
\includegraphics[width=5cm]{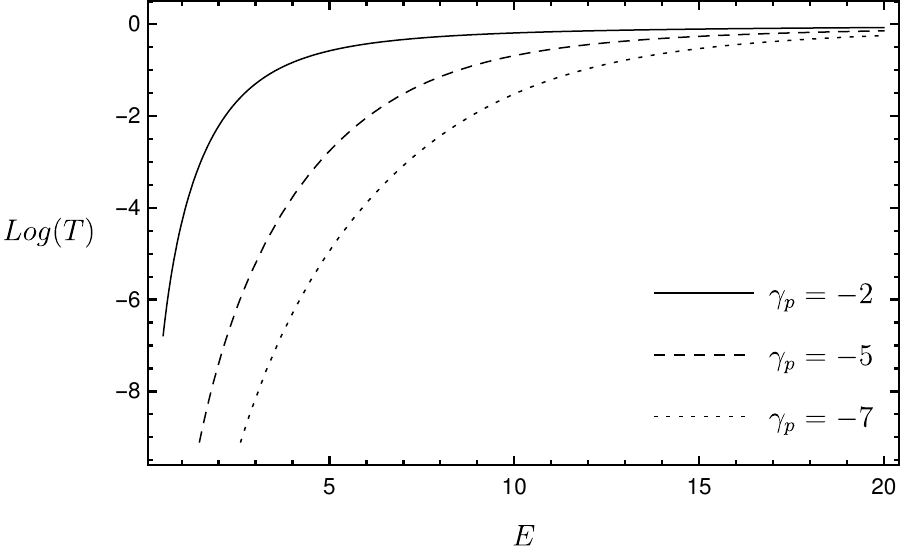}
\label{fig:TKR-R4-RS}
}
\caption{Transmission 
coefficient for KR field in Randall-Sundrum delta like  scenario with $k=1$ as 
function of  $E = m_{T}^{2}$. (a) Ricci, Einstein and Horndeski tensor coupling,
(b) Riemann tensor coupling.}
\end{figure}

For the massive case, the Eq. (\ref{eqKR2}) provides the solution
\begin{equation}\label{psimasRRS}
 \psi(z) =(k|z|+1)^{1/2}[C_{1}J_{\nu}(m_{T}|z|+ m_{T}/k)+C_{2}Y_{\nu}(m_{T}|z|+ 
m_{T}/k)],
\end{equation}
where $C_{1}$ and $C_{2}$ are constants with $\nu=\sigma+1/2$ for Ricci, Einstein and Horndeski tensors coupling and $\nu=\sqrt{-2\gamma}$ for Riemann tensor coupling. 
Since the Bessel functions goes to 
infinity as $(m_{T}|z|+ m_{T}/k)^{-1/2}$, no fixation of constants $C_{1}$ and 
$C_{2}$ produces a convergent solution.
 Then the massive modes are non-localized. To obtain more information
about massive modes we can evaluate the transmission coefficient. For this, we 
will write the solution (\ref{psimasRRS}) in the form
\begin{equation}
 \psi(z) = \left\lbrace\begin{matrix}E_\nu(-z)+r F_\nu(-z) &,\;\mbox{for}\; z<0 \\ 
t F_\nu(z)&,\;\mbox{for}\; z\geq0\end{matrix}\right.,
\end{equation}
where
\begin{eqnarray}
&&E_\nu(z) = \sqrt{\frac{\pi}{2}}(m_{T}z+ m_{T}/k)^{1/2}H^{(2)}_{\nu}(m_{T}z+ 
m_{T}/k),
\\&& F_\nu(z) = \sqrt{\frac{\pi}{2}}(m_{T}z+m_{T}/k)^{1/2}H^{(1)}_{\nu}(m_{T}z+ 
m_{T}/k),
\end{eqnarray}
 $H^{(1)}_{\nu}$ and $H^{(2)}_{\nu}$ are the Hankel functions of first and second 
kind respectively, $r$ and $t$ are constants. The boundary conditions at $z = 
0$ 
imposes 
\begin{equation}
 t =  \frac{W(E_\nu,F_\nu)(0)}{ 2F_\nu(0)F_\nu'(0) + 2k\beta F_\nu^{2}(0)},
\end{equation}
where $ W(E_\nu,F_\nu)(0) =  E_\nu(0)F_\nu'(0)-E_\nu'(0)F_\nu(0)$ is the Wronskian  at $z=0$. Since the 
Wronskian is constant in Schr\"odinger equation, the transmission coefficient 
can be written as
\begin{equation}
T = |t|^{2} = \frac{m_{T}^{2}}{|F_\nu(0)F_\nu'(0) + k\beta  F_\nu^{2}(0)|^{2}}. 
\end{equation}
The transmission coefficient was plotted in Fig. \ref{fig:TREH-RS-D5p2} as 
function of $E=m_T^2$ for the KR field in the Ricci, Einstein and Horndeski coupling and does not show peaks, indicating no unstable massive modes. 
The transmission coefficient was plotted for Riemann coupling in Fig. \ref{fig:TKR-R4-RS} and does not show peaks, indicating no unstable massive modes. 

For the vector field  in Randall-Sundrum scenario the potential of 
Schr\"odinger equation, (\ref{potvec2}), can be written as
\begin{equation}\label{potvecRRS}
U(z)=\frac{(\alpha+\beta)k^{2}}{(k|z|+1)^{2}} -2k\beta\delta(z).
\end{equation}
This is the same potential of KR filed, so the behavior of the modes of vector 
field is the same, i.e., the zero mode is localized while the massive are not. 
The transmission coefficient for the massive modes is the same of KR field, Figs. 
\ref{fig:TREH-RS-D5p2} and \ref{fig:TKR-R4-RS}.

\subsection{In brane scenario generated by a domain-wall}
 In this section we will use the smooth warp factor produced by a 
domain-wall\cite{Du:2013bx,Melfo:2002wd},
\begin{equation}\label{WF-dw}
  A(z) = -\frac{1}{2n}\ln\left[\left(kz\right)^{2n}+1\right],
\end{equation}
which recover the Randall-Sundrum metric at large $z$ for $n \in N^{*}$. 

\begin{figure}[!h]
\centering
\includegraphics[width=5cm]{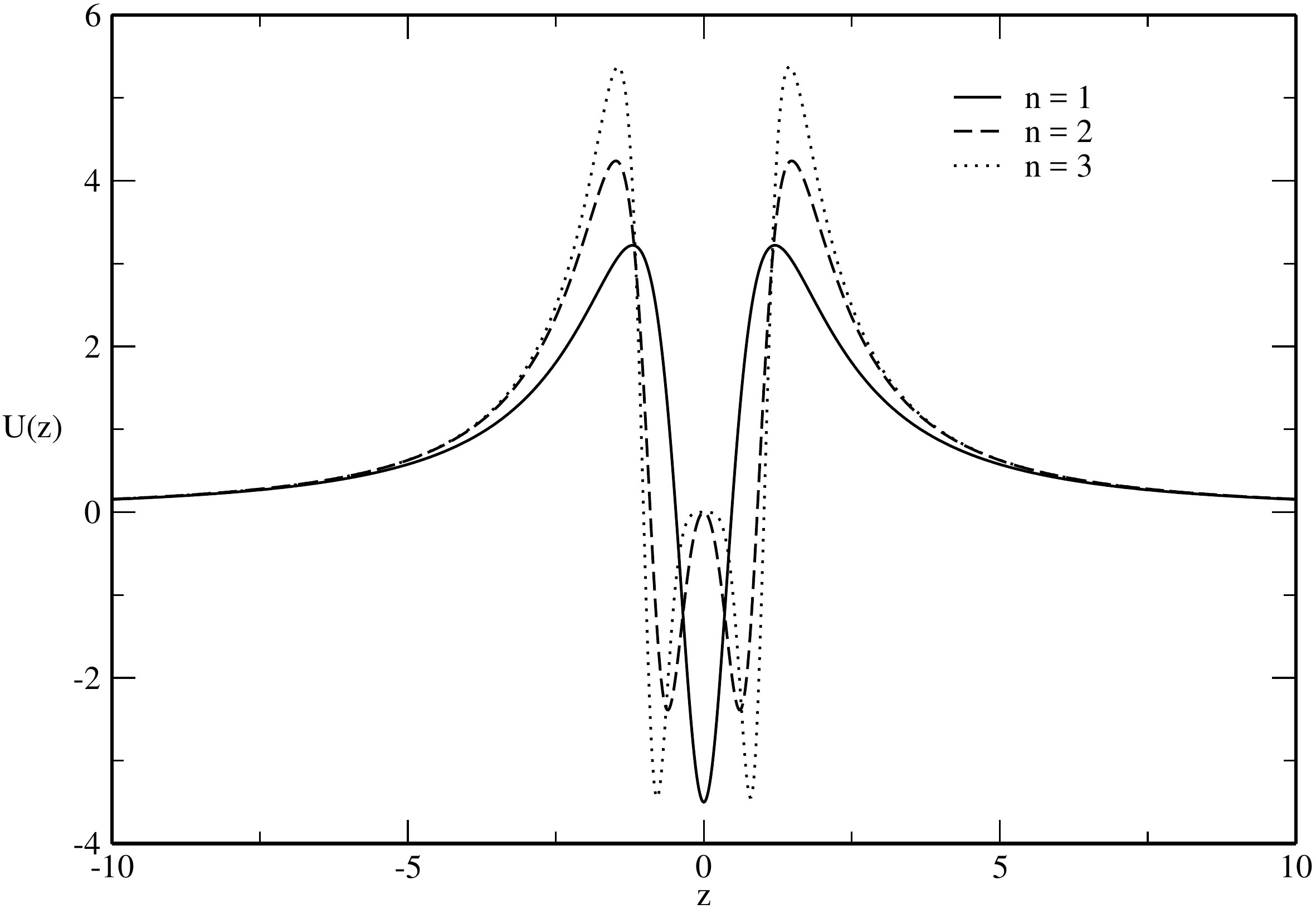}
\caption{Schr\"odinger's potential for KR field in smooth scenario generated by domain 
walls for Ricci tensor coupling with some values of parameter $n$ and $k=1$.}
\label{fig:potKR-Rmn-sm}
\end{figure}

Using this metric in Eq. (\ref{potKR2}) we obtain the Schr\"odinger's potential 
for transversal part of KR field
\begin{equation}\label{potRsm}
U_T(z)=\frac{(k z)^{2 n} \left(\alpha (k z)^{2 n}+\beta \left((k z)^{2 n}-2 n+1\right)\right)}{z^2 \left((k z)^{2 n}+1\right)^2},
\end{equation}
which is illustrated in Fig. \ref{fig:potKR-Rmn-sm}, for Ricci tensor coupling with some values of $n$. 

\begin{figure}[!h]
\centering
\subfigure[]{
\includegraphics[width=5cm]{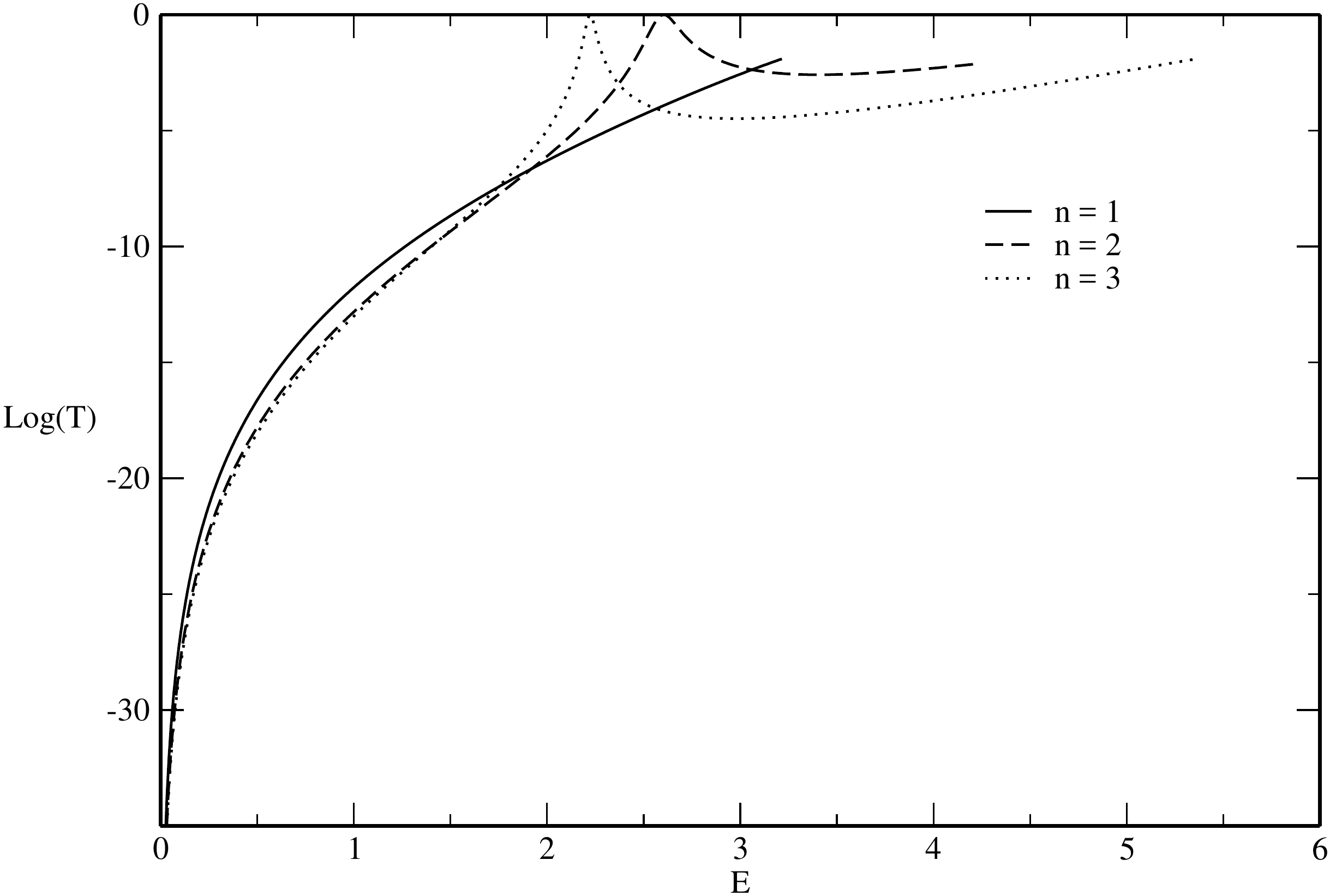}
\label{fig:TKR-Rmn-sm}
}
\subfigure[]{
\includegraphics[width=5.5cm]{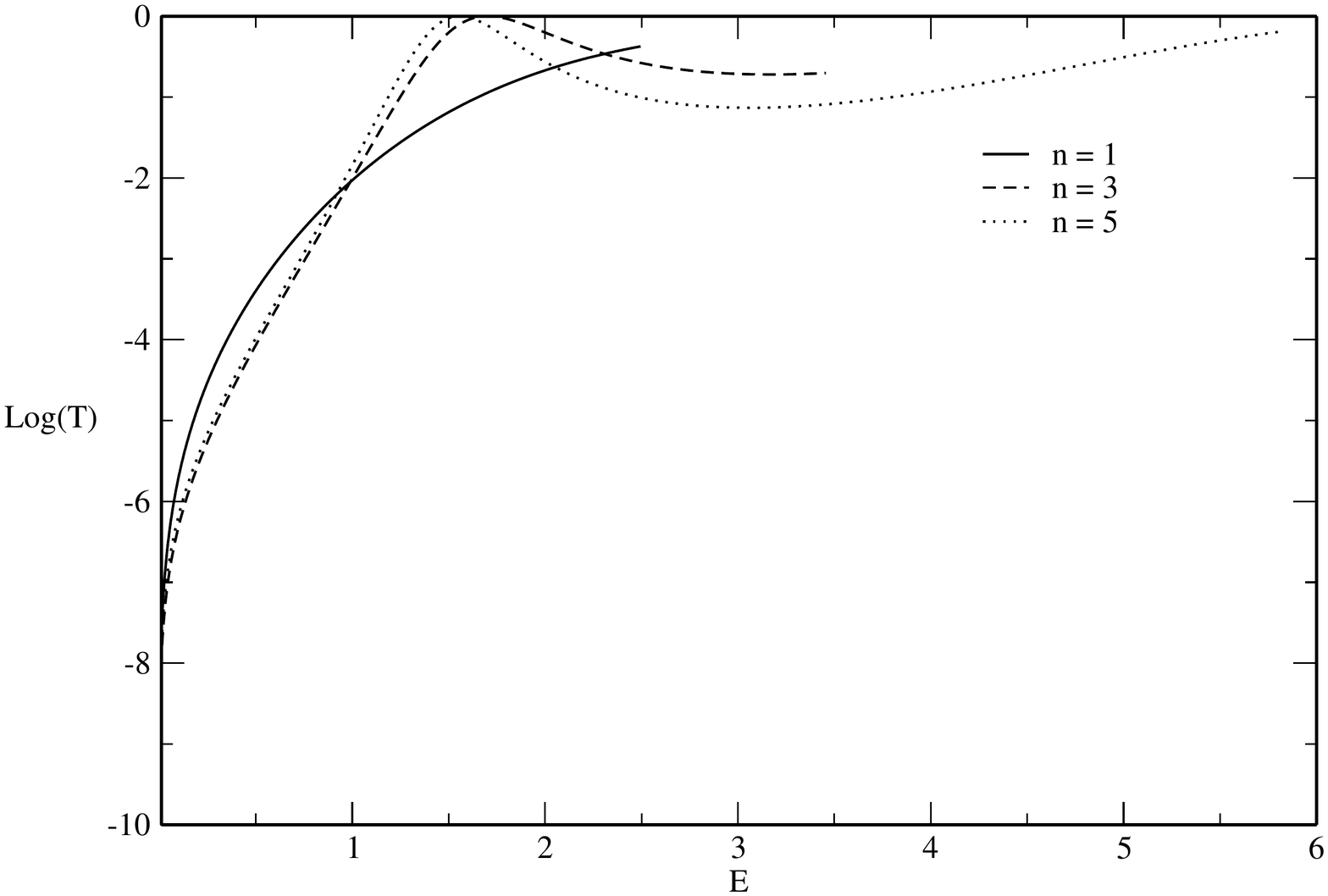}
\label{fig:TGmn-smn}
}
\subfigure[]{
\includegraphics[width=5.5cm]{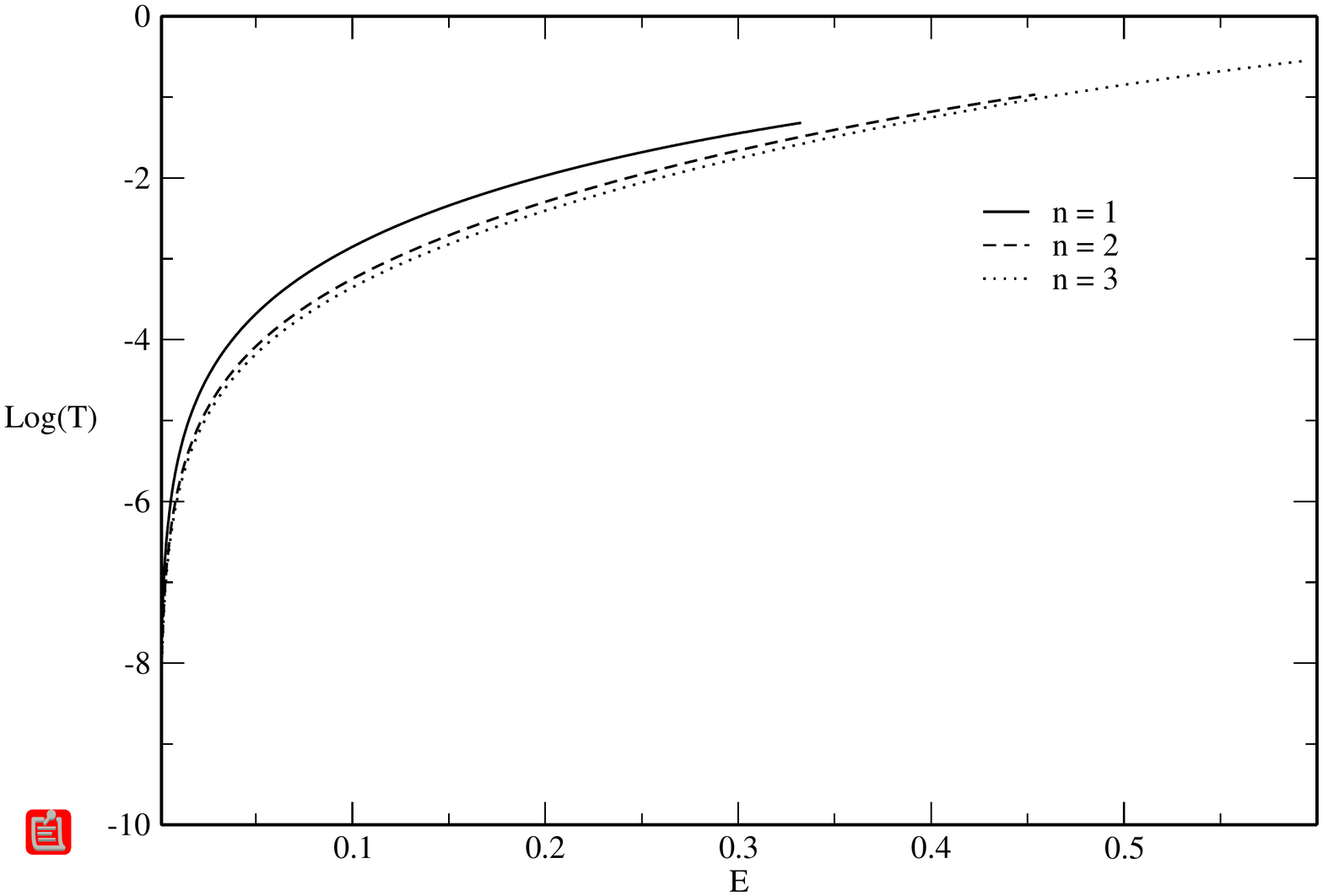}
\label{fig:TKR-D4-sm}
}
\caption{The transmission 
coefficient in a smooth scenario generated by domain walls  for some values of 
parameter $n$ as function of  $E= m_{T}^{2}$. (a) Ricci tensor, (b) Einstein tensor and  (c) Horndeski tensor.}
\end{figure}

The solution of massive modes of  transversal of KB field can not be found 
analytically. To obtain information about this state we use the transfer matrix 
method to evaluate the transmission coefficient. 
The behavior of the transmission coefficient for Ricci, Einstein and Horndeski coupling is illustrated in Figs. \ref{fig:TKR-Rmn-sm}, \ref{fig:TGmn-smn} and \ref{fig:TKR-D4-sm}  for some values of 
parameter $n$. As we can see, for Ricci tensor coupling, resonant peaks appears when we increase the 
values of the parameter $n$ indicating the  existence of unstable massive modes. The same occur for Einstein tensor coupling. In the Horndeski coupling 
we observe the absence of resonant peaks.

\begin{figure}[!h]
\centering
\subfigure{
\includegraphics[width=5cm]{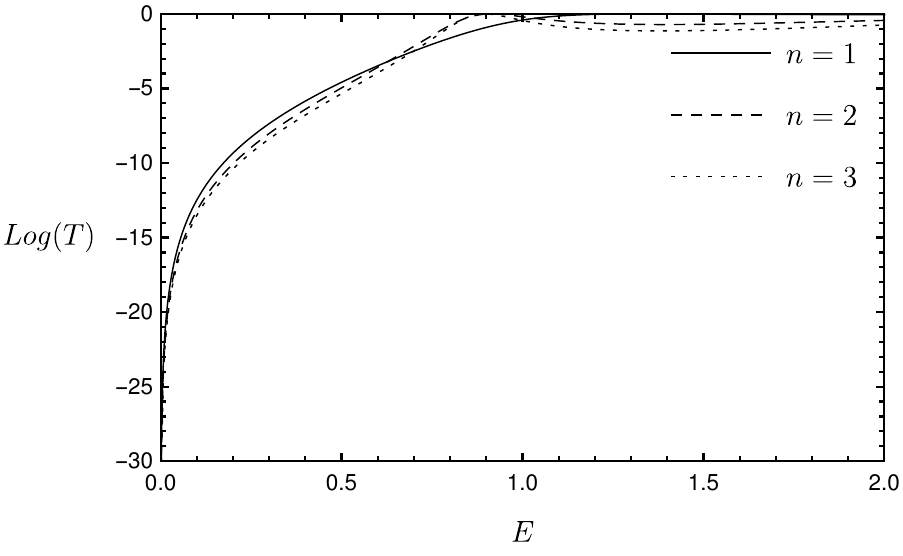}
\label{fig:TR4-smn}
}
\subfigure{
\includegraphics[width=5cm]{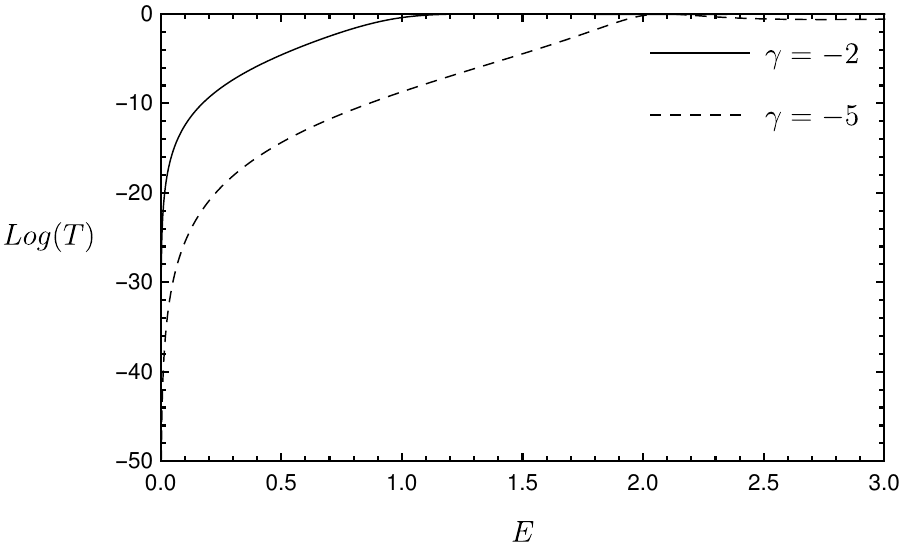}
\label{fig:TGmn-smg25}
}
\caption{Transmission coefficient for KR field in brane scenario generated 
by domain-walls with Riemann tensor coupling as a function of  $E = 
m_{T}^{2}$. (a) For some values of $n$ with $\gamma = -2$. (b) For some values 
of coupling constant $\gamma$ with $n=1$.}
\end{figure}

The behavior of the transmission coefficient for Riemann coupling  is illustrated in Fig. \ref{fig:TR4-smn} for some values of 
parameter $n$ with $\gamma = -2$ and in Fig. \ref{fig:TGmn-smg25} for some 
values of coupling constant $\gamma$ with $n =1$. As we can see, when we increase the values of 
$n$ and $|\gamma|$ we observe the appearance of resonant peaks, indicating the 
existence of unstable massive modes.

\begin{figure}[!h]
\centering
\subfigure[]{
\includegraphics[width=5cm]{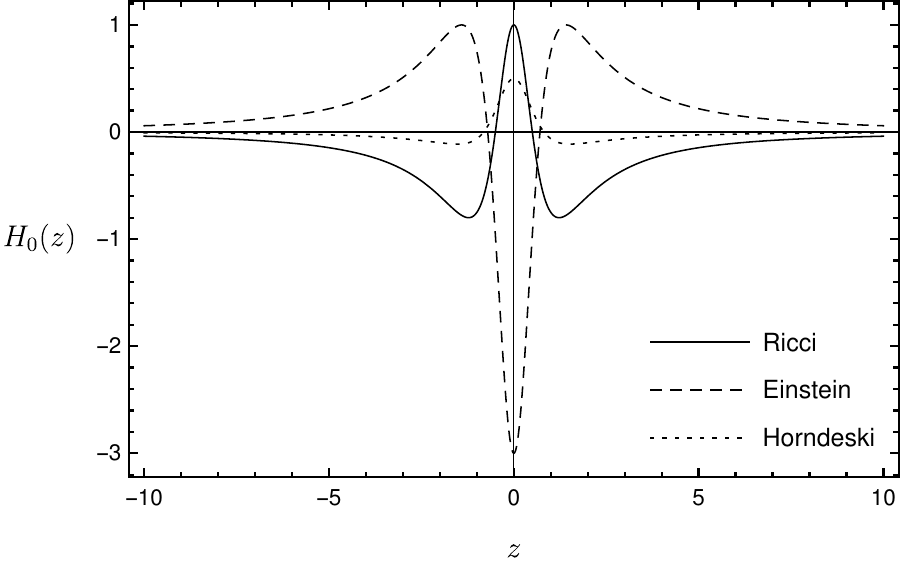}
\label{fig:ProfileH0-REH-DW}
}
\subfigure[]{
\includegraphics[width=5cm]{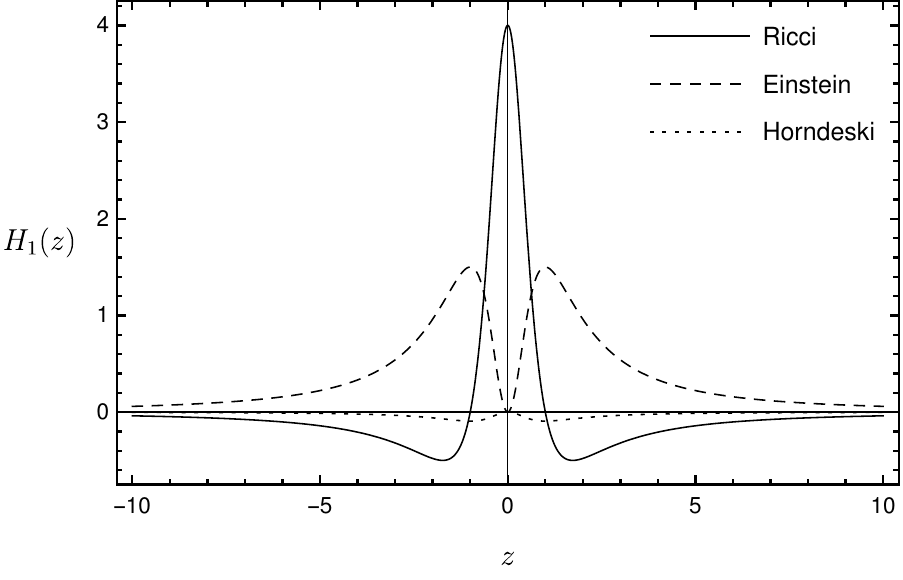}
\label{fig:ProfileH1-REH-DW}
}
\caption{The components $H_0$ and $H_1$ in smooth scenario generated by domain 
wall with $n=1$ for the Ricci, Einstein and Horndeski tensors. (a) $H_0$ and (b) $H_1$.}
\end{figure}

For the reduced vector field the potential is given by Eq. (\ref{potvec2}).
The components $H_0$ and $H_1$ vanishes at regular points near to the origin 
for all values of parameter $n$ and the potential diverges at this 
same points (see Figs. \ref{fig:ProfileH0-REH-DW} and \ref{fig:ProfileH1-REH-DW}). These kind of divergence does 
not allow us to use the transfer matrix method to compute the transmission 
coefficient and to evaluate the existence of unstable massive modes.

\subsection{In brane scenario generated by a kink} 
\begin{figure}[!h]
\centering
\includegraphics[width=5cm]{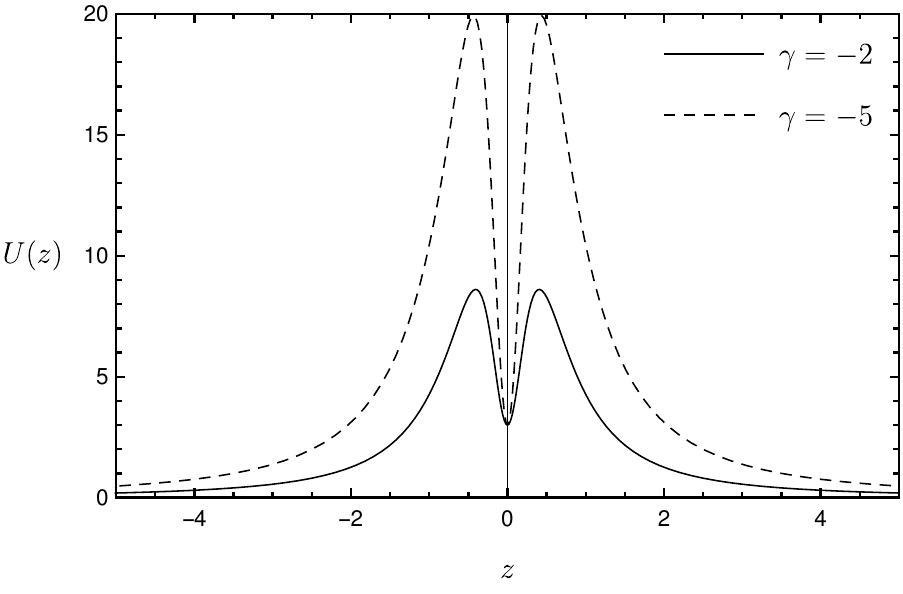}
\caption{(a) Schr\"odinger potential in kink scenario for KR field coupled to 
Riemann tensor for some values of coupling constant $\gamma$.}
\label{fig:potKR-R4-kink}
\end{figure}

\begin{figure}[h]
\centering
\subfigure[]{
\includegraphics[width=5cm]{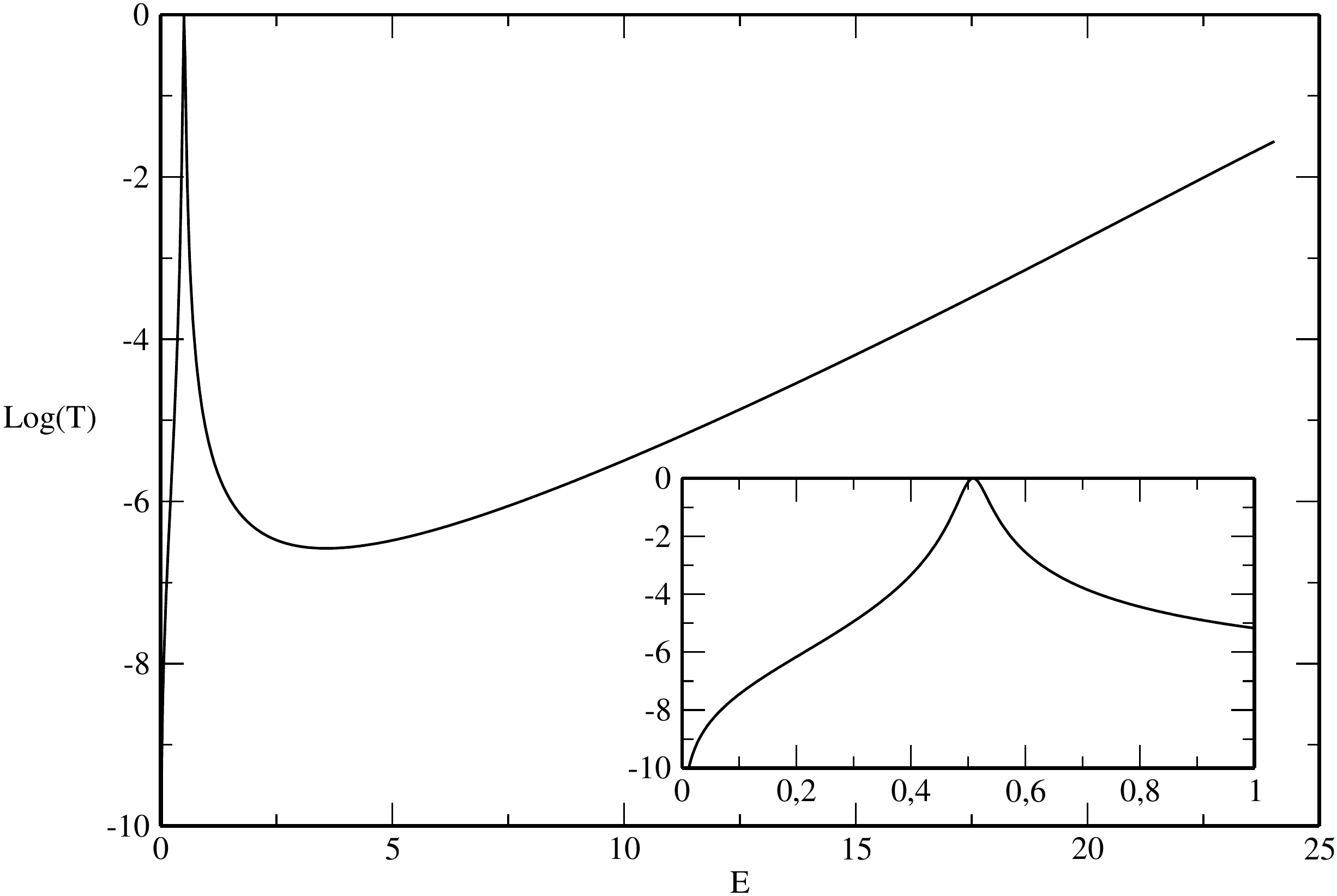}
\label{fig:TKR-Rmn-kink}
}
\subfigure[]{
\includegraphics[width=5.5cm]{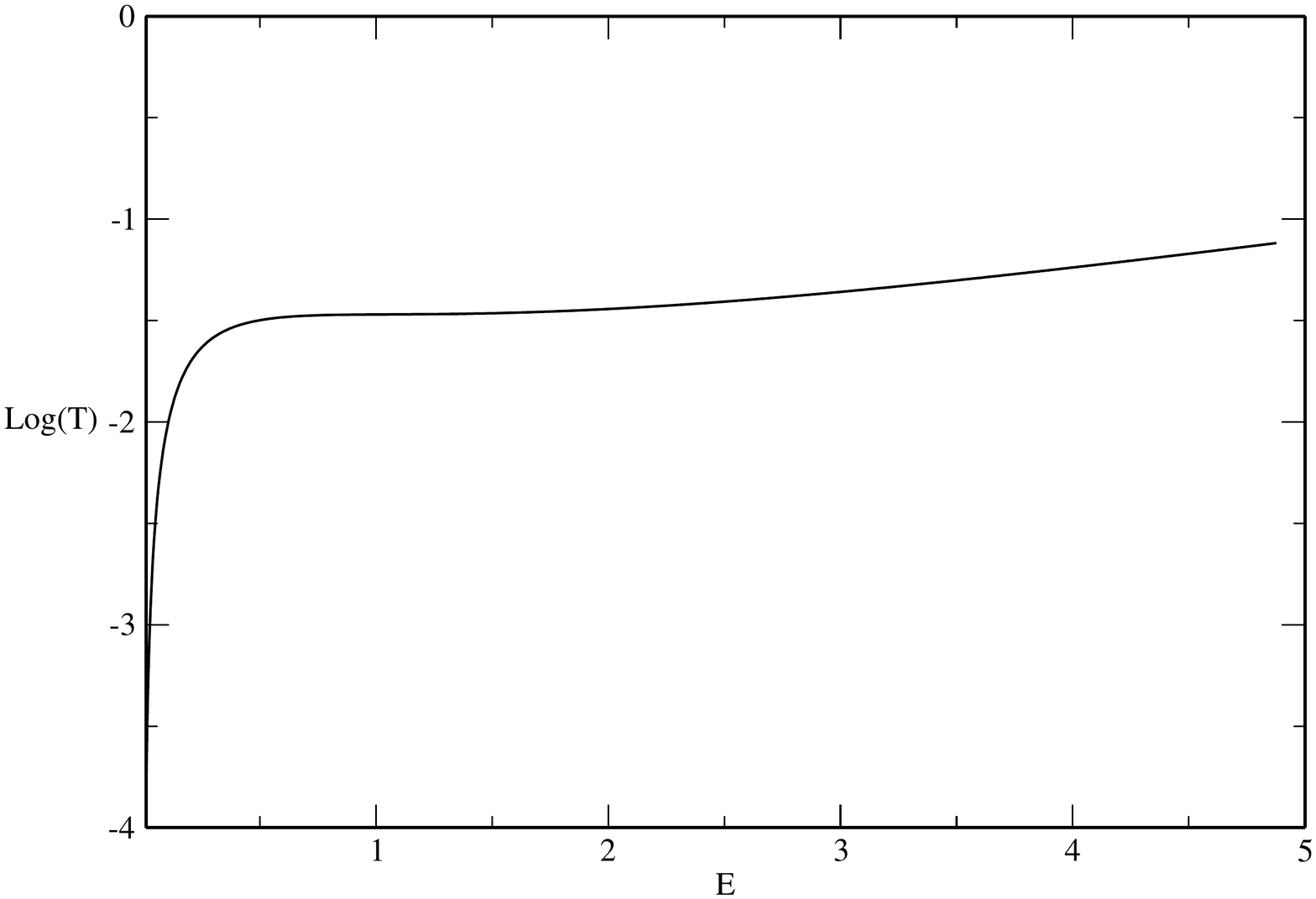}
\label{fig:TKR-Gmn-kink}
}
\caption{Transmission coefficient in kink scenario for KR field as 
function of $E= m_{T}^{2}$. (a) Ricci coupling and (b) Einstein coupling.}
\end{figure}

\begin{figure}[!h]
\centering
\subfigure[]{
\includegraphics[width=5cm]{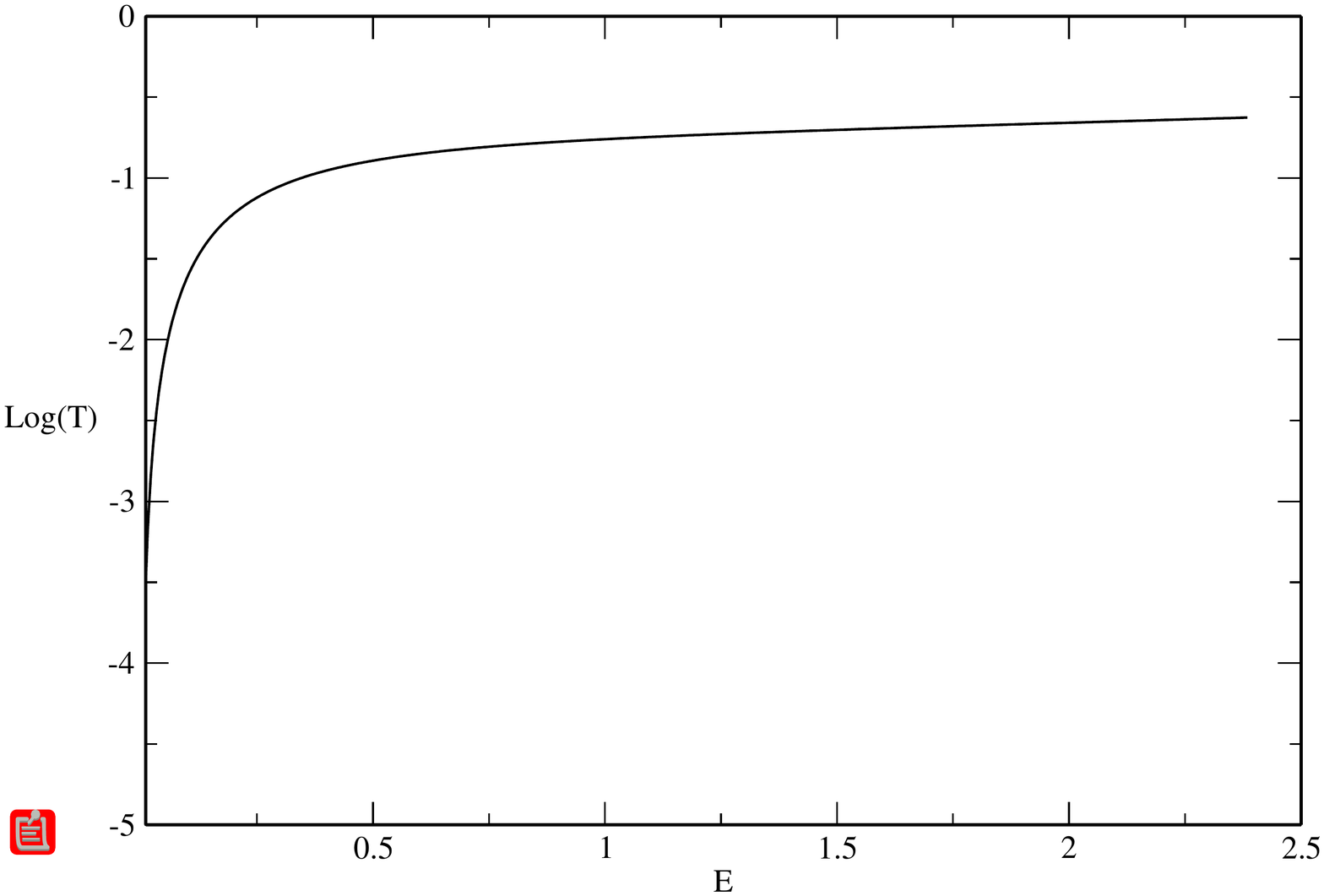}
\label{fig:TKR-D4-kink}
}
\subfigure[]{
\includegraphics[width=5cm]{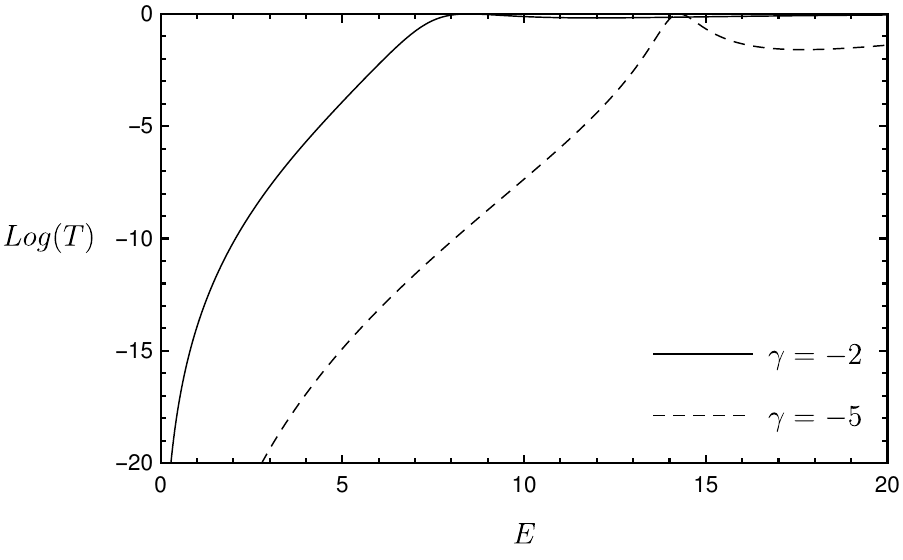}
\label{fig:TKR-R4-kink}
}
\caption{Transmission coefficient in kink scenario for KR 
field as function of $E= m_{T}^{2}$. (a) Horndeski coupling
 and (b) Riemann coupling}
\end{figure}

For a four dimensional brane generated by a kink, the warp factor is given 
by\cite{Landim:2011ki}
\begin{equation}\label{WF-kink}
 A(y) = -4\ln\cosh y -\tanh^{2}y,
\end{equation}
where the variable $y$ relates with the conformal coordinate, $z$, by
\begin{equation}
dz = \e^ {-A(y)}dy.
\end{equation}
 The behavior of potential of transversal part of KB field, Eq. 
(\ref{potKR2}), with this warp factor is illustrated in Fig. \ref{fig:potKR-R4-kink}
 for Riemann coupling. For massive modes, like in 
previous sections, we use the matrix transfer method to compute the transmission coefficient. The 
result is plotted in Figs. \ref{fig:TKR-Rmn-kink}, \ref{fig:TKR-Gmn-kink}, \ref{fig:TKR-D4-kink} and \ref{fig:TKR-R4-kink}. As we can see, for Ricci tensor coupling  we have a resonant peak near $E=0.5$ and near $E=14$ with $\lambda=-5$ for Riemann coupling.

 Like the previous case, the components $H_0$ and $H_1$  vanishes at regular 
points near to the origin and  the potential diverges at this  points. These 
kind of divergence does not allow us to use the transfer matrix method to 
compute the transmission coefficient of vector field and to evaluate the 
existence of unstable massive modes.

\section{The $p-$form zero mode case}
In this section we further develop the previous methods in order to generalize 
our results to the $p-$form field case in a $(D-1)$-brane. We again must 
consider the coupling to tensors of order two and four.

\subsection{The $p-$form coupled with a rank two geometric tensor}
In this subsection we will consider the coupling of the $p-$form field to rank 
two geometric tensors. The action is given by
\begin{equation}
S_{p}=-\frac{1}{2p!}\int 
d^{D}x\sqrt{-g}\left[\frac{(Y_{M_{1}...M_{p+1}})^2}{(p+1)!}+\gamma_{p}g_{M_1N_1}
H^{M_{1}M_{2}}X_{M_{2}M_{3}...M_{p+1}}X^{N_{1}M_{3}...M_{p+1}})\right],
\end{equation}
where $Y_{M_{1}...M_{p+1}}=\partial_{[M_{1}}X_{M_{2}...M_{p+1}]}$. The 
equations 
of motion are given by
\begin{equation}\label{H2motionpform}
\frac{1}{p!} 
\partial_{M_{1}}[\sqrt{-g}Y^{M_{1}...M_{p+1}}]-\frac{1}{2}\sqrt{-g}\gamma_{p}g_{
M_1N_1}H^{M_{1}[M_{2}}X^{N_{1}M_{3}]...M_{p+1}}=0. 
 \end{equation}
Similarly to the KR case, from the above equation we get the constraint
\begin{equation}\label{H2divpform} 
\partial_{M_{2}}\left(g_{M_1N_1}H^{M_{1}[M_{2}}\sqrt{-g}X^{N_{1}M_{3}]...M_{p+1}
}\right)=0.
\end{equation}

Now we must decompose the $p-$form in $D$-dimensions to a $p-$form and a 
$(p-1)-$form in $(D-1)$-dimensions. For this we must expand Eq.  
(\ref{H2motionpform}) and use Eq.(\ref{H2}).
We arrive at just two kinds of terms: one where none of the
indexes are $D-1$ and another where one of the indexes is $D-1$
\begin{eqnarray}
&&\frac{1}{p!}\e^{\alpha_{p}A}\partial_{\mu_{1}}[Y^{\mu_{1}\mu_{2}...\mu_{p+1}}]
+\frac{1}{p!}\partial_z(\e^{\alpha_{p}A}Y^{D-1\,\mu_{2}...\mu_{p+1}})\nonumber\\
&&-\gamma_{p} H_0\e^{\alpha_{p}A}X^{\mu_{2}...\mu_{p+1}}=0,\\
&&\frac{1}{p!}\partial_{\mu_{1}}Y^{\mu_{1}\mu_{2}...\mu_{p}\,D-1} 
-\frac{H_0+H_1}{2}\gamma_{p}X^{\mu_{2}...\mu_{p}\,D-1}=0,
\end{eqnarray}
with $\alpha_{p}=D-2(p+1)$.  We should point that for the one form case the 
second of the above formulas is not valid. This is due to the fact that in this 
case the $H$ tensor will not be anti-symmetrized with any index of the field. 
In 
fact for this case the equation is given by
\begin{equation}
\partial_{\mu_{1}}Y^{\mu_{1}\,D-1} -H_1\gamma_{p}X^{D-1}=0
\end{equation}
and the the vector field, in principle, should be considered separately. 
However 
we can unify both equations in the following way
\begin{eqnarray}
&&\frac{1}{p!}\e^{\alpha_{p}A}\partial_{\mu_{1}}[Y^{\mu_{1}\mu_{2}...\mu_{p+1}}]
+\frac{1}{p!}\partial_z(\e^{\alpha_{p}A}Y^{D-1\,\mu_{2}...\mu_{p+1}})\nonumber\\
&&-\gamma_{p} H_0\e^{\alpha_{p}A}X^{\mu_{2}...\mu_{p+1}}=0,\label{H2pformnu}\\
&&\frac{1}{p!}\partial_{\mu_{1}}Y^{\mu_{1}\mu_{2}...\mu_{p}\,D-1} -\frac{\kappa 
H_0+H_1}{\kappa+1}\gamma_{p}X^{\mu_{2}...\mu_{p}\,D-1}=0,\label{H2pform5}
\end{eqnarray}
where $\kappa=0$ for the gauge field and $\kappa=1$ in the other cases. In the 
next section we will see that this will provide a powerful simplification of 
the problem.

Just as in the KR case, the equation (\ref{H2divpform}) gives rise to two 
equations. For one free index equals to $D-1$ we get 
$\partial_{\mu_{1}}X^{\mu_{1}...\mu_{p-1}\,D-1} \equiv
\partial_{\mu_{1}}X^{\mu_{1}...\mu_{p-1}}=0$, where we have used our previous 
definitions. Therefore we see null divergence condition for our $(p-1)-$form 
field is naturally obtained upon
dimensional reduction. For all index different of $D-1$ we get 
\begin{equation}\label{H2transversepform}
\partial_z\left[\frac{\kappa 
H_0+H_1}{\kappa+1}\e^{\alpha_{p}A}X^{\mu_{1}...\mu_{p-1}}\right]+ 
H_0\e^{\alpha_{p}A}\partial_{\mu_{p}}X^{\mu_{1}...\mu_{p}}=0.
\end{equation}

The above equation says to us that the $p-$form does not has null divergence. 
In order to obtain a consistent zero mode over the brane we must decouple the 
longitudinal and transversal parts defined by
\begin{eqnarray}\label{pformlong}
X_{T}^{\mu_{1}...\mu_{p}}\equiv 
X^{\mu_{1}...\mu_{p}}+\frac{(-1)^{p}}{\Box}\partial^{[\mu_{1}}\partial_{\nu_{1}}
X^{\mu_{2}...\mu_{p}]\nu_{1}},\\
X_{L}^{\mu_{1}...\mu_{p}}\equiv 
\frac{(-1)^{p-1}}{\Box}\partial^{[\mu_{1}}\partial_{\nu_{1}}X^{\mu_{2}...\mu_{p}
]\nu_{1}},
\end{eqnarray}
from where we get
\begin{equation}
\partial_{\mu_{1}}Y^{\mu_{1}\mu_{2}...\mu_{p+1}}=p!\square 
X_{T}^{\mu_{2}...\mu_{p+1}},\;\;\partial_{\mu_{1}}Y^{\mu_{1}\mu_{2}...\mu_{p}}
=p!\square X_{T}^{\mu_{2}...\mu_{p}},
\end{equation}
and
\begin{equation}
Y^{D-1\,\mu_{1}...\mu_{p}}=Y_{L}^{D-1\,\mu_{1}...\mu_{p}}+p! X_{T}^{'\mu_{1}...\mu_{p}}.
\end{equation}

With these we can write the equations (\ref{H2pformnu}) and (\ref{H2pform5}) as
\begin{eqnarray}
&&\e^{\alpha_{p}A}\square 
X_{T}^{\mu_{1}...\mu_{p}}+\partial_z(\e^{\alpha_{p}A}\partial_z 
X_{T}^{\mu_{1}...\mu_{p}})-\gamma_{p}H_0\e^{\alpha_{p}A}X_{T}^{\mu_{2}...\mu_{
p+1}}\nonumber\\
&&+\frac{1}{p!}\partial_z(\e^{\alpha_{p}A}Y_{L}^{D-1\,\mu_{1}...\mu_{p}})-\gamma_{p}
H_0\e^{\alpha_{p}A}X_{L}^{\mu_{1}...\mu_{p}}=0,\label{H2eqXTXLpform}\\
&&\frac{1}{p!}\partial_{\mu_{1}}Y_{L}^{\mu_{1}\mu_{2}...\mu_{p}}-\gamma_{p}\frac
{\kappa H_0+H_1}{\kappa+1}X^{\mu_{2}...\mu_{p}}=0\label{H2eqYLphipform}. 
\end{eqnarray}

Therefore, we see clearly from Eq.  (\ref{H2eqXTXLpform}) that we have a 
coupling between
the transversal part of the $p-$form field, the longitudinal part and the 
$(p-1)-$form
field. From Eq.  (\ref{H2eqYLphipform}) we see that the $(p-1)-$form is coupled 
to the longitudinal part of the $p-$form field. As in the case of the KR field, we 
should expect that we have to uncouple the effective massive equations for the 
gauge fields $X_{T}^{\mu_{1}\mu_{2}...\mu_{p}}$ and $X^{\mu_{2}...\mu_{p}}$ 
since both satisfy the null divergence condition in $(D-1)$ dimensions. Lets 
walk along and prove this now. First of all note that using 
$\partial_{\mu_2}X^{\mu_{2}...\mu_{p}}=0$
we can show that 
\begin{equation}\label{H2pforminverse}
Y_L^{\mu_{1}...\mu_{p}}=\frac{(-1)^{p-1}}{\Box}\partial^{[\mu_{1}}\partial_{\nu}
Y_L^{\mu_{2}...\mu_{p}]\nu}
\end{equation}
and using equations (\ref{H2eqYLphipform}) and (\ref{H2transversepform}) we 
obtain
\begin{equation}
\partial_z\left(\e^{\alpha_{p}A}Y_{L}^{\mu_{1}...\mu_{p}\,D-1}\right)=
p!\gamma_{p}H_0\e^{\alpha_{p}A}X_{L}^{\mu_{1}...\mu_{p}}. 
\end{equation}
From this we can see that the longitudinal component in Eq. 
(\ref{H2eqXTXLpform}) decouples. For the $(p-1)-$form field we just  use 
(\ref{H2transversepform}) in equation (\ref{H2eqYLphipform}). Then we get for 
the decoupled equations of motion
\begin{eqnarray}
&& \e^{\alpha_{p}A}\square 
X_{T}^{\mu_{1}...\mu_{p}}+\partial_z(\e^{\alpha_{p}A}\partial_z 
X_{T}^{\mu_{1}...\mu_{p}})-\gamma_{p}H_0\e^{\alpha_{p}A}X_{T}^{\mu_{1}...\mu_{p}
}=0,\label{XTpfull}\\
&&\square 
X^{\mu_{2}...\mu_{p}}+\partial_z\left[\frac{\e^{-\alpha_p}}{H_0}\partial_z\left[
\frac{\kappa 
H_0+H_1}{\kappa+1}\e^{\alpha_{p}A}X^{\mu_{2}...\mu_{p}}\right]\right]\nonumber\\
&&-\gamma_{p}\frac{\kappa H_0+H_1}{\kappa+1}X^{\mu_{2}...\mu_{p}}=0.
\end{eqnarray}

To study the mass spectrum of the $p-$form we must impose the separation of 
variables in the form $ X_{T}^{\mu_{1}...\mu_{p}}(z,x) = 
\tilde{X}_{T}^{\mu_{1}...\mu_{p}}(x)\e^{-\alpha_{p}A/2}\psi(z)$ to obtain 
\begin{eqnarray}
&&(\square-m^2) \tilde{X}_{T}^{\mu_{1}...\mu_{p}}=0,\\
&&\psi''-U_T(z)\psi =-m^2\psi,
\end{eqnarray}
where
\begin{equation}\label{H2potp}
U_T(z)= \frac{\alpha_{p}^{2}}{4}A'^{2}+\frac{\alpha_{p}}{2}A''+\gamma_pH_0.
\end{equation}

To the $(p-1)-$form field case we must separate the variables in the form 
$X^{\mu_{2}...\mu_{p}}= \tilde{X}^{\mu_{2}...\mu_{p}}F(z)\psi(z)$, where
\begin{equation}\label{H2Fpform}
 F(z) = \e^{\alpha_pA/2}({H}_{0})^{1/2}\frac{\kappa+1}{\kappa H_0+H_1}.
\end{equation}
This procedure provides the following set of equations
\begin{eqnarray}
&& (\Box-m^2)\tilde{X}^{\mu_{2}...\mu_{p}}= 0,
\\&& \psi''-U(z)\psi  = -m^{2}{H}_{0}\frac{\kappa+1}{\kappa H_0+H_1}\psi, 
\label{eq(p-1)2}
\end{eqnarray}
where the potential of the Schr\"odinger equation is given by
\begin{equation}\label{pot(p-1)2}
 U(z) = \frac{1}{4}\left[\alpha_pA' +(\ln{H}_{0})'\right]^{2} 
-\frac{1}{2}\left[\alpha_pA +\ln{H}_{0}\right]'' +\gamma{H}_{0}
\end{equation}
but we should be careful, since for this case the conditions for localizability 
is that
\begin{equation}\label{integrand(p-1)}
\int \frac{H_{0}}{(\kappa H_{0}+H_{1})^2}\psi^2dz,
\end{equation}
is finite. In the next section we will see that we can find a master equation 
governing the trapping of fields. Therefore we stop here our analyzes.

\subsection{The $p-$form coupled with a rank four geometric tensor}
Finally we must study the last case of the manuscript, namely the coupling of 
the $p-$form field to rank four geometric tensors. Since we have already 
constructed and defined all the relevant quantities before this will be very 
direct. The action is given by
\begin{equation}
S_{p}=-\frac{1}{2p!}\int 
d^{D}x\sqrt{-g}\left[\frac{(Y_{M_{1}...M_{p+1}})^2}{(p+1)!}+\gamma_{p}H^{N_{1}N_
{2}}_{N_{3}N_{4}}X_{N_{1}N_{2}M_{3}...M_{p+1}}X^{N_{3}N_{4}M_{3}...M_{p+1}}
)\right],
\end{equation}
and the equations of motion are given by
\begin{equation}\label{H4motionpform}
\frac{1}{p!} 
\partial_{M_{1}}[\sqrt{-g}Y^{M_{1}...M_{p+1}}]-\frac{1}{2}\sqrt{-g}\gamma_{p}H_{
N_{1}N_{2}}^{[M_2M_3}X^{N_{1}N_{2}M_{4}]...M_{p+1}}=0. 
 \end{equation}

Again, due to the anti-symmetry, we get the constraint
\begin{equation}\label{H4divpform} 
\partial_{M_{2}}\left(\sqrt{-g}H_{N_{1}N_{2}}^{[M_2M_3}X^{N_{1}N_{2}M_{4}]...M_{
p+1}}\right)=0.
\end{equation}

Following the same steps as before, by using Eqs.(\ref{H4motionpform}) and 
Eq.(\ref{H4}), we obtain the coupled equations of motion for the $p-$form and 
$(p-1)-$forms in $(D-1)$-dimensions
\begin{eqnarray}
&&\frac{1}{p!}\e^{\alpha_{p}A}\partial_{\mu_{1}}[Y^{\mu_{1}\mu_{2}...\mu_{p+1}}]
+\frac{1}{p!}\partial_z(\e^{\alpha_{p}A}Y^{D-1\,\mu_{2}...\mu_{p+1}})\nonumber\\
&&-\gamma_{p} H_0\e^{\alpha_{p}A}X^{\mu_{2}...\mu_{p+1}}=0,\label{H4pformnu}\\
&&\frac{1}{p!}\partial_{\mu_{1}}Y^{\mu_{1}\mu_{2}...\mu_{p}\,D-1} -\frac{\kappa 
H_0+H_1}{\kappa+1}\gamma_{p}X^{\mu_{2}...\mu_{p}\,D-1}=0,\label{H4pform5}
\end{eqnarray}
where now $\kappa=0$ for $p=2$ and $\kappa=1$ for the other cases. Since 
$\kappa$ depends on the degree of the form and on the rank of the tensor, from 
now on we will use $\kappa_{p,r}$. In this notation we have that 
$\kappa_{p,r}=0$ for the pairs $(p,r)=(1,2)$ and $(p,r)=(2,4)$. For the 
constraint we get again from the component $D-1$ of   
(\ref{H4divpform}) $\partial_{\mu_{1}}X^{\mu_{1}...\mu_{p-1}}=0$ and for all 
index different of $D-1$ we get 
\begin{equation}\label{H4transversepform}
\partial_z\left[\frac{\kappa_{p,r} 
H_0+H_1}{\kappa_{p,r}+1}\e^{\alpha_{p}A}X^{\mu_{1}...\mu_{p-1}}\right]+ 
H_0\e^{\alpha_{p}A}\partial_{\mu_{p}}X^{\mu_{1}...\mu_{p}}=0.
\end{equation}

As said before, the use of the parameter $\kappa_{p,r}$ provides a powerful way 
to simplify and unify the problem of localization for $p-$form fields. As we 
can see Eqs. (\ref{H4pformnu}),(\ref{H4pform5}) and (\ref{H4transversepform}) are 
identical to Eqs. (\ref{H2pformnu}),(\ref{H2pform5}) and 
(\ref{H2transversepform}). Therefore this can be seem as a master equation that 
governs the $p-$form fields non-minimally coupled to gravity in RS scenarios. 
By separating the variables we also arrive at the following master equations that 
drives the spectrum of the reduced $p-$form and $(p-1)-$form fields.

\begin{eqnarray}
&&(\square-m^2) \tilde{X}_{T}^{\mu_{1}...\mu_{p}}=0\label{mastereqp}\\
&&\psi''-U_T(z)\psi =-m^2\psi, \\
&&U_T(z)= \frac{\alpha_{p}^{2}}{4}A'^{2}+\frac{\alpha_{p}}{2}A''+\gamma_pH_0, 
\label{masterpotp}
\end{eqnarray}
and
\begin{eqnarray}
&&(\Box-m^2)\tilde{X}^{\mu_{2}...\mu_{p}}= 0, \label{mastereq(p-1)}
\\ &&\psi''-U(z)\psi  = -m^{2}{H}_{0}\frac{\kappa_{p,r}+1}{\kappa_{p,r} 
H_0+H_1}\psi,\\
&&U(z) = \frac{1}{4}\left[\alpha_pA' +(\ln{H}_{0})'\right]^{2} 
-\frac{1}{2}\left[\alpha_pA +\ln{H}_{0}\right]'' +\gamma{H}_{0}.
\end{eqnarray}

The analyzes is identical to the one performed in the second section. To the 
$p-$form we find that the field is localized if $\lambda_0\neq 0$ and
\begin{equation}\label{pcond}
p>\frac{(D-2)}{2}-\frac{\beta_0}{\lambda_0},
\end{equation}
with
\begin{equation}\label{gammapD}
\gamma_p=\frac{\beta_0-\lambda_0\alpha_p}{\lambda^2_0}.
\end{equation}
For the $(p-1)-$form we have
\begin{equation}\label{(p-1)cond}
p<\frac{D}{2}+\frac{\beta_0}{\lambda_0}, 
\end{equation}
with
\begin{equation}\label{gamma(p-1)D}
\gamma_p=\frac{\beta_0+\lambda_0(\alpha_p+2)}{\lambda^2_0}.
\end{equation}

In the conclusion section we must discuss the general consequences of the above results. However we should stress the fact that simple formulas can be obtained 
depending only in few parameters for all cases considered. Now we must test if some unstable massive modes can be found over the brane 
with the above couplings.

\section{The $p-$form massive modes}
As we saw in the last section, the transverse potential for Ricci, Einstein 
and Horndeski tensors couplings  have the form
\begin{equation}\label{UTPform}
 U_T(z)=\sigma_p^2A'(z)^2+\sigma_p A''(z),
\end{equation}
with localizable zero mode if $\sigma_p>0$, where 
$\sigma_p=-D/2+p+\beta_0/\lambda_0+1$. Is important to note that 
$\sigma_p=p-1/2,p-1$ for the Einstein, Horndeski tensor respectively. This 
implies that the possible unstable massives modes does not depend of dimension 
of the space in these cases. 

The  Schr\"odinger transverse potential  when  the  $p-$form field is coupled 
to the Riemann tensor has the form
\begin{equation}\label{UTPform-Riemann}
U_T(z)=\left(\frac{\alpha_p^2}{4}-2\gamma_p\right)A'^2+\frac{\alpha_p}{2}A'',
\end{equation}
where $\gamma_p<(\alpha_p^2+2\alpha_p)/8$, to have positive asymptotic behavior.

In the following sections we study the possible 
existence of  unstable massive modes for the Ricci, Einstein, Horndeski  and Riemann tensors 
couplings in Randall-Sundrum delta like, domain wall and kink scenarios.
\subsection{In Randall-Sundrum delta like scenario}
\begin{figure}[!h]
\centering
\includegraphics[width=5cm]{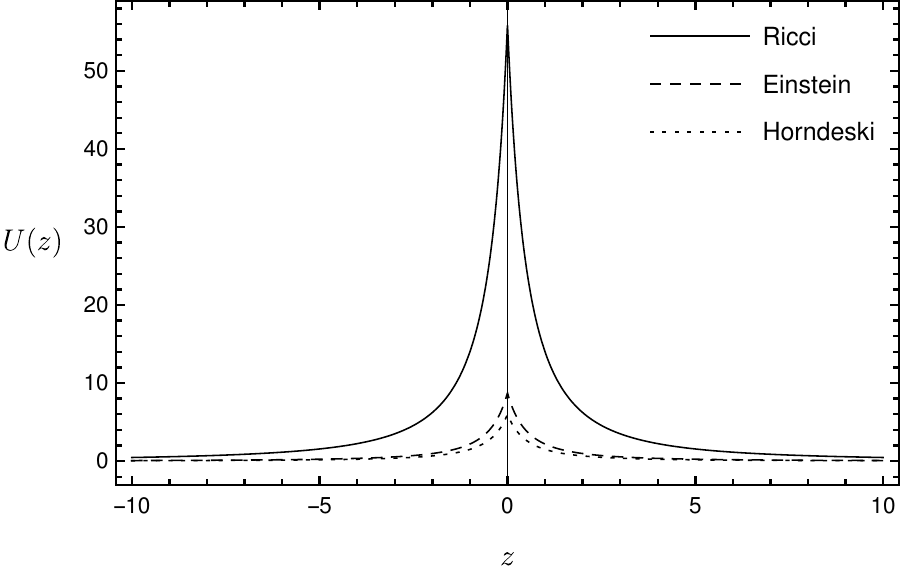}
\caption{(a) Regular part of Schr\"odinger potential for a $p$-form field in 
Randall-Sundrum delta like scenario with $k =1, D=10$ and $p=3$ for Ricci, 
Einstein and Horndeski couplings.}
 \label{fig:ProfileREH-RS-D10p3}
\end{figure}

\begin{figure}[!h]
\centering
\subfigure[]{
\includegraphics[width=5cm]{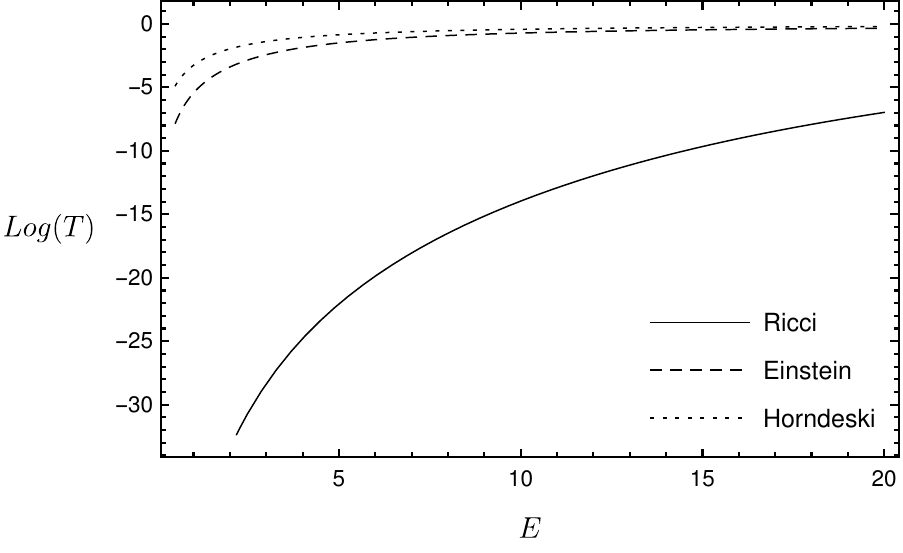}
\label{fig:TREH-RS-D10p3}
}
\subfigure[]{
\includegraphics[width=4.9cm]{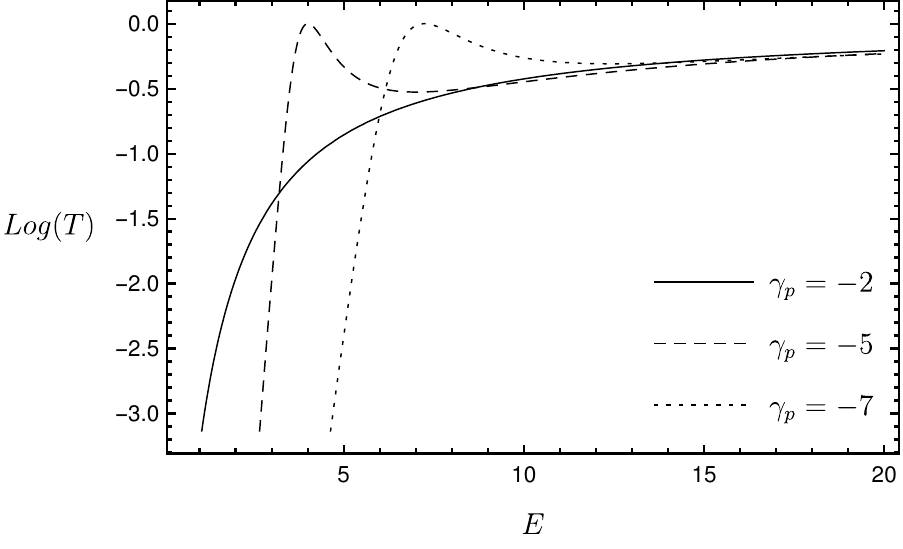}
\label{fig:T-Riemann-RS-D10p3}
}
\caption{(a) The  transmission coefficient in Randall-Sundrum delta like scenario with $k =1, D=10$ and $p=3$ as 
function of  $E = m_{T}^{2}$. (a) Ricci, Einstein, Horndeski and (b) Riemann.}
\end{figure}

In this section we will use the warp factor given by Eq. (\ref{WF-delta}).
For the transversal part of the $p$-form field, the potential of Schr\"odinger equation, 
Eq. (\ref{UTPform}), is given by 
\begin{equation}\label{potR2RS-p}
U_T(z)=\frac{k^2\sigma_p(\sigma_p+1)}{(k|z|+1)^{2}} -2k\sigma_p\delta(z).
\end{equation}
By making  the same procedure of the sections above, we obtain the transmission 
coefficient
\begin{equation}\label{TDP-1}
T = |t|^{2} = \frac{m_{T}^{2}}{|F_{\nu_p}(0)F'_{\nu_p}(0) + k\sigma_p  
F^{2}_{\nu_p}(0)|^{2}},
\end{equation}
with $\nu_p=|\sigma_p+1/2|$ and 
$F_{\nu_p}(z)=\sqrt{\frac{\pi}{2}}(m_{T}z+m_{T}/k)^{1/2}H^{(1)}_{\nu_p}(m_{T}z+ 
m_{T}/k)$.
We show in Fig. \ref{fig:ProfileREH-RS-D10p3}  the regular part of the 
Schr\"odinger potential in Randall-Sundrum delta like scenario for Ricci, 
Einstein and Horndeski couplings. As we can see the Ricci tensor gives a bigger 
maximum for the potential which means that only 
larger masses pass through the brane. We hope that only resonant peaks appears 
for large masses for Ricci tensor coupling. As we can see in  Fig. 
\ref{fig:TREH-RS-D10p3}, no resonant peaks appear in the Randall-Sundrum 
scenario for the Ricci, Einstein and Horndeski tensor coupling.

For the Riemann coupling the transversal part of $p$-form field, the potential of Schr\"odinger equation, 
Eq. (\ref{UTPform-Riemann}), is given by 
\begin{equation}\label{potR2RS-Riemann-p}
U_T(z)=\frac{k^2(\alpha_p^2+2\alpha_p-8\gamma_p)}{4(k|z|+1)^2} 
-k\alpha_p\delta(z).
\end{equation}
  The transmission coefficient in this case is given by
\begin{equation}\label{TDP-2}
T = |t|^{2} = \frac{4m_{T}^{2}}{|2F_{\nu_p}(0)F'_{\nu_p}(0) + k\alpha_p  
F^{2}_{\nu_p}(0)|^{2}},
\end{equation}
with $\nu_p=\sqrt{(\alpha_p+1)^2-8\gamma_p}/2$ and 
$F_{\nu_p}(z)=\sqrt{\frac{\pi}{2}}(m_{T}z+m_{T}/k)^{1/2}H^{(1)}_{\nu_p}(m_{T}z+ 
m_{T}/k)$.
The maximum of Schr\"odinger potential  grows as  $|\gamma_p|$ increases.  As we can see in  Fig. 
\ref{fig:T-Riemann-RS-D10p3}, resonant  peaks appear in the Randall-Sundrum 
delta like scenario for the Riemann tensor coupling when $|\gamma_p|$ increases.

\subsection{In brane scenario generated by a domain-wall}
\begin{figure}[!h]
\centering
\subfigure[]{
\includegraphics[width=4.9cm]{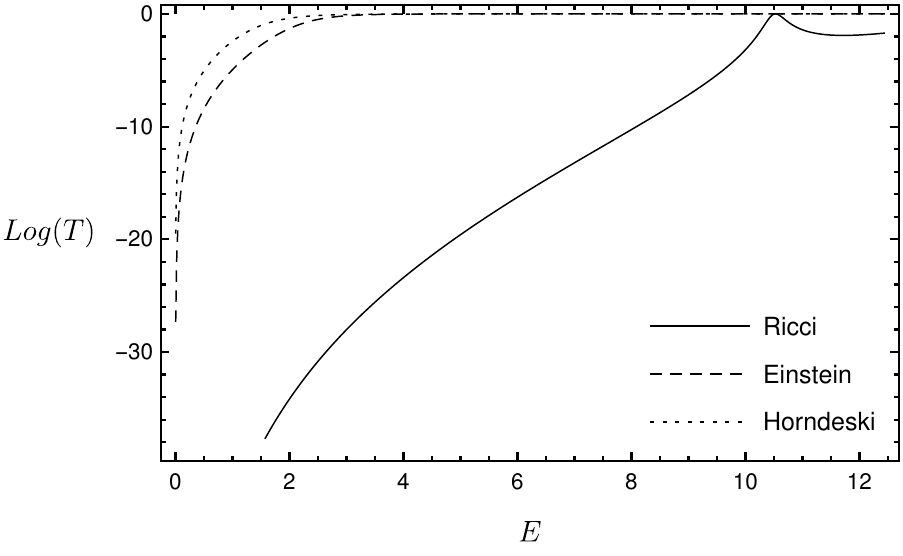}
\label{fig:TREH-DW-D10p3n1}
}
\subfigure[]{
\includegraphics[width=4.9cm]{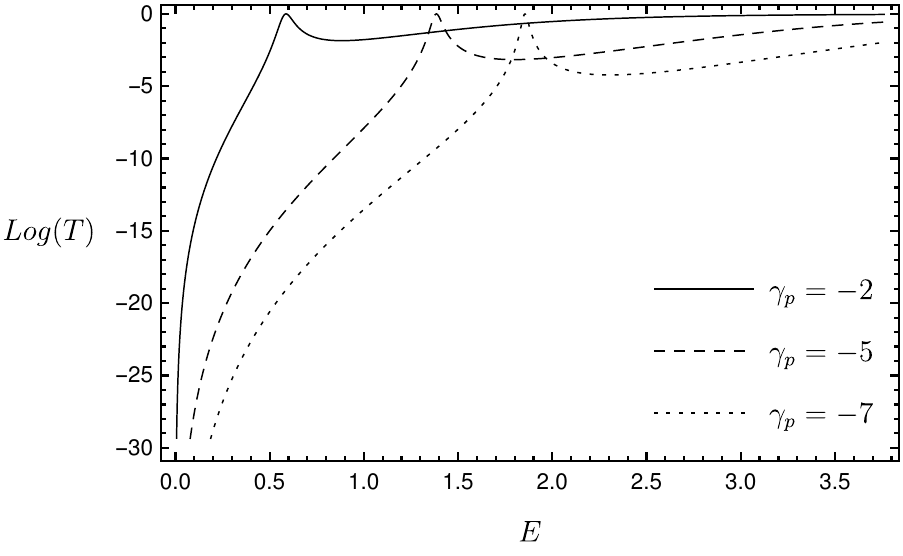}
\label{fig:T-Riemann-DW-D10p3n1}
}
\caption{The transmission coefficient  for a $p-$form  field in a Domain-wall 
with $n=1,D=10$ and $p=3$ as function of  $E = m_{T}^{2}$. (a) Ricci, Einstein, Horndeski and (b) Riemann. }
\end{figure}

Now we analyze all tensor coupling in a brane 
scenario generated a domain-wall. As in the Kalb-Ramond case, the warp factor 
used will be given by Eq. (\ref{WF-dw}).
As we can see in  Fig. \ref{fig:TREH-DW-D10p3n1}, no resonant peaks appear in 
Einstein and Horndeski coupling. However we have a resonant peak  $m_T^2=10.5$ 
for the Ricci tensor coupling. For the Riemann coupling, in  Fig. \ref{fig:T-Riemann-DW-D10p3n1}, we observe three 
resonant peaks.

\subsection{In brane scenario generated by a kink}
\begin{figure}[!h]
\centering
\subfigure[]{
\includegraphics[width=5cm]{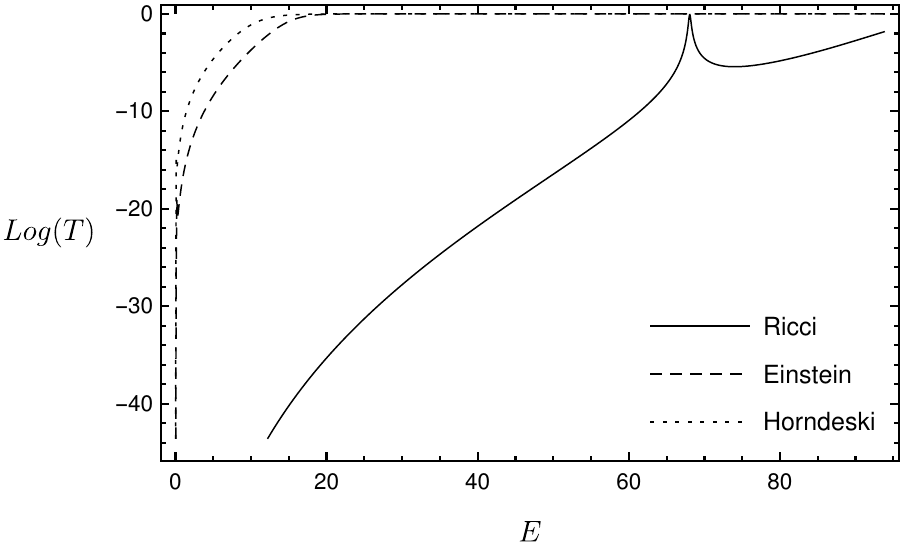}
\label{fig:TREH-KINK-D10p3}
}
\subfigure[]{
\includegraphics[width=5cm]{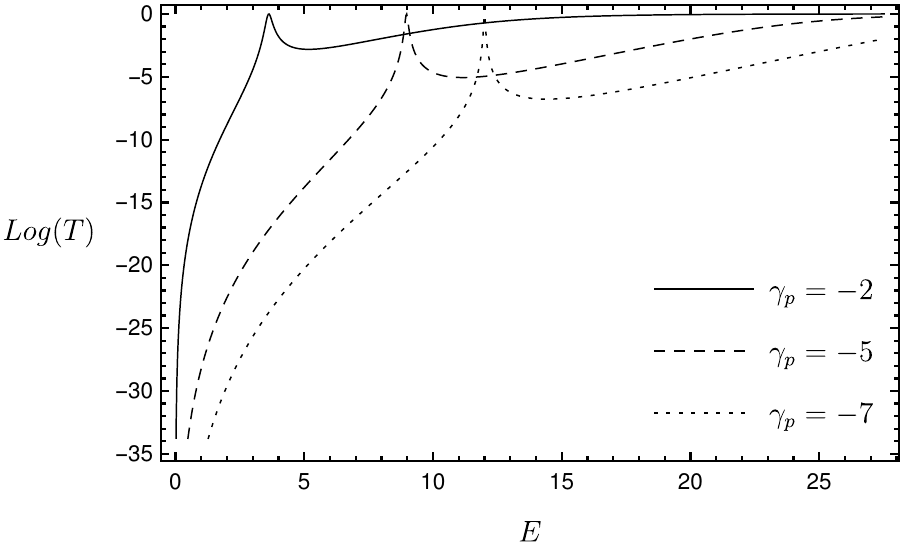}
\label{fig:T-Riemann-KINK-D10p3}
}
\caption{(a) The transmission coefficient for a $p-$form  field in a brane 
scenario generated by a kink with $D=10$ and $p=3$ as function 
of  $E = m_{T}^{2}$. (a) For Ricci, Einstein and 
Horndeski couplings. (b) For Riemann coupling.}
\end{figure}

Here we analyze all tensor coupling in a brane 
scenario generated by a kink. As in the Kalb-Ramond case the warp factor used 
will be given by Eq. (\ref{WF-kink}). As we can see in  Fig. \ref{fig:TREH-KINK-D10p3}, no resonant peaks appear in 
Einstein and Horndeski coupling again. However we have a resonant peak  
$m_T^2=67$ for the Ricci tensor coupling. From  Fig. 
\ref{fig:T-Riemann-KINK-D10p3}, we can see that resonant peaks appear in the Riemann coupling again. 

\section{Conclusion}
In this paper we analyzed the localization of $p-$form field in co-dimension 
one  brane scenarios non-minimally coupled to gravity. First we consider the 
Kalb-Ramond field coupled to rank two and four geometric tensors. We show that 
the reduced fields can be decoupled in a similar way as in the vector field 
case. 
We analyze the localization of zero mode of the transversal part of KR field 
for a generic geometric tensor and found the 
conditions to localize it. For the vector component of the KR field, the study 
of localization is more complicated, due to the potential of the Schr\"odinger 
equation and the analise could be done only asymptotically. We find that for 
both, the reduce KR and vector component, the fields are localized for the 
Ricci, Einstein and Horndeski tensors but not for the Riemann tensor. We find 
that the value of the coupling constant is the same for both and therefore consistent.  
To analyze the localizability for general $p$, we use the values of 
$\beta_0/\lambda_0$ and substitute in Eqs. (\ref{pcond}) and (\ref{(p-1)cond}) 
to obtain Table \ref{tab1}.
\begin{table}[ht]
\begin{center}
\begin{tabular}{|l|c|c|c|}\hline 
{\bf Tensor}  & {\bf$\beta_0/\lambda_0$}& 
{\bf$p>$}& {\bf$p<$} \\ \hline
Einstein &  $(D-3)/2$ & $1/2$& $(2D-3)/2$\\ \hline
Horndeski &  $(D-4)/2$ & $1$& $(D-2)$ \\ \hline
Ricci  &  $(D-2)$& $-(D-2)/2$& $(3D-4)/2$\\ \hline
Riemann &  $-$ & $-$& $-$\\ \hline
\end{tabular}
\caption{The localizability condition for the $p-$forms fields.}\label{tab1}
\end{center}
\end{table}
From this we can see that for some values of the parameters both components of 
the reduced $p-$form can be localized. For example, for the Ricci tensor and 
for 
$D=5$ we have that the reduced $p-$form is localized for $p>-1$ and therefore 
for any value of $p$. For the the $(p-1)-$form in the same case the condition 
is 
given by $p<11/2$, what means all the cases, since in five dimensions the 
bigger 
value of $p$ is five. However beyond this there is a second consistence 
condition. The coupling constants (\ref{gammapD}) and (\ref{gamma(p-1)D}) must 
be the same. By imposing that $\gamma_p=\gamma_{(p-1)}$ we find that 
$p=(D-1)/2$. Therefore the condition for having both components localized is 
universal and independent of the kind of coupling used and just depends on the 
dimension of spacetime. In five dimensions for example the Kalb-Ramond field 
generates a Kalb-Ramond plus a vector field trapped over the membrane for any 
kind of coupling. Another important result of the Table \ref{tab1} is that, as said 
in the introduction, we can test some previous hypothesis of previous works. In 
Ref. \cite{Alencar:2015oka} the coupling of the vector field with the Ricci and 
the Einstein tensor has been studied. It has been found that for the second 
case 
it is not possible to localize the field. The hypothesis was that tensors with 
zero divergence do not provides a localized field. However from the above table 
we can see that the Einstein tensor can trap any $p-$form in any dimension with 
only one exception: the case $p=1$. Therefore somehow the hypothesis is right 
but is only valid for the vector field. The Ricci tensor also can trap any 
$p-$form in any $D$ with one exception: the gauge field in $D=2$. The Riemann 
tensor can not localize any field. The Horndeski tensor can trap any field 
since this coupling is possible for $p>1$ and the localization condition is given by 
$p>3/2$.  

Now we will consider massive modes. This is done for all 
geometric tensors  for many kinds of smooth branes. 
In the case of massives modes, 
we used the transmission coefficient to observe possible unstable massive 
modes. The emergence of resonant peaks was observed when we increased the 
coefficients 
of $A''(z)$ and $A'(z)$. From Figs. (\ref{fig:TREH-RS-D5p2}) and (\ref{fig:TKR-R4-RS}) we observed the absence of 
resonance for KR field in RS delta like branes for all tensor coupling. The same occur for $p$-forms field in $D=10$, for Einstein, Horndeski and Ricci coupling as 
we can see from Fig. (\ref{fig:TREH-RS-D10p3}). In delta like brane, we observe the appearance of resonance only in Riemann coupling 
in $D=10$ (see Fig. (\ref{fig:T-Riemann-RS-D10p3}). For domain wall branes, for KR field, we concluded from 
Figs. (\ref{fig:TKR-Rmn-sm}), (\ref{fig:TGmn-smn}), (\ref{fig:TKR-D4-sm}), (\ref{fig:TR4-smn}) and  (\ref{fig:TGmn-smg25}), the resonance 
appear only for Ricci and Riemann coupling with increasing value of parameter $n$. The same occur for $p$-forms field in $D=10$. The conclusion is the same for kink branes. 
This can be explaining  as follows.
As we can see from Eq. (\ref{UTPform}), when we 
increase $\sigma_p$, $A'(z)$ predominates over $A''(z)$. The same occur for $\gamma_p<0$ in Eq. (\ref{UTPform-Riemann}). The behavior of $U(z)$ looks 
like a double barrier, for domain wall and kink like branes (see the Figs. (\ref{fig:potKR-Rmn-sm}) and 
(\ref{fig:potKR-R4-kink})). When we increase the values of the parameters, the width of the barrier increase and we have more probability to see resonant peaks.
For the domain wall brane, for large $n$,  $A'(z)\sim |z|^{-2}$ for $|z|>1$ and $A'(z) \sim 0$ for $|z|<1$. That is, for large $n$  
and higger dimension, we have a Schr\"odinger potential like   a double delta barrier with two deltas located at $z=\pm1$. As pointed in the section 6, for the Einstein and 
Horndeski coupling, the Schr\"odinger potential does not depend on dimension of 
space. Consequently, the resonant peaks will appear for greater values of the 
form $p$ in Einstein and Horndeski coupling. Since in the Ricci and Riemann tensor 
coupling, the potential depends on the dimension of space-time, it was observed, 
that these cases  are more sensitive to the presence of unstable massive modes as showed in Table \ref{tab2}.

\begin{table}[ht]
\begin{center}
\begin{tabular}{|l|c|c|c|}\hline 
{\bf Tensor} & {\bf Delta like}  & {\bf Domain Wall} & {\bf Kink}  \\ \hline
Einstein & - & - & -\\ \hline
Horndeski & - & - & -\\ \hline
Ricci  & - & + & + \\ \hline
Riemann & + & +& +\\ \hline
\end{tabular}
\caption{The plus signal means appearance of resonant peaks, the minus absence for $p-$forms.}\label{tab2}
\end{center}
\end{table}

\section*{Acknowledgments}

The authors would like to thanks Alexandra Elbakyan and sci-hub, for removing all barriers in the way of science.
We acknowledge the financial support provided by Funda\c c\~ao Cearense de 
Apoio ao Desenvolvimento Cient\'\i fico e Tecnol\'ogico (FUNCAP)  through PRONEM PNE-0112-00085.01.00/16 and the Conselho 
Nacional de Desenvolvimento Cient\'\i fico e Tecnol\'ogico (CNPq).

\end{document}